\documentclass[12pt,preprint]{aastex}

\begin{document}

\setcounter{equation}{0}
\setcounter{figure}{0}

\def\thebib{\section*{}\list
 {[\arabic{enumi}]}{\settowidth\labelwidth{[]}\leftmargin\labelwidth
 \advance\leftmargin\labelsep
 \usecounter{enumi}}
 \def\newblock{\hskip .11em plus .33em minus .07em}
 \sloppy\clubpenalty4000\widowpenalty4000
 \sfcode`\.=1000\relax}
\let\endthebib=\endlist
\def\etal{{et al.}}
\def\apj{ApJ}
\def\apjs{ApJ Supp.}
\def\apjl{ApJ Lett.}

\def\ltsima{$\; \buildrel < \over \sim \;$}
\def\simlt{\lower.5ex\hbox{\ltsima}}
\def\gtsima{$\; \buildrel > \over \sim \;$}
\def\simgt{\lower.5ex\hbox{\gtsima}}
\def\hmpc{\;h^{-1}{\rm Mpc}}
\def\kms{{\rm \;km\;s^{-1}}}

\def\lya{{\rm Ly}$\alpha$\ }


\def\ni{\noindent}        
\def\ub{\underbar}
\def\hi{\noindent \hangindent=2.5em}
\def\et{{\it et\thinspace al.}}    
\def\pc{{\rm\,pc}}
\def\kpc{{\rm\,kpc}}
\def\Mpc{{\rm\,Mpc}}
\def\mpc{{\rm\,Mpc}}
\def\hnot{{\rm\,km/s/Mpc}}
\def\kmsec{{\rm\,km/s}}
\def\kms{{\rm\,km/s}}
\def\msun{{\rm\,M_\odot}}
\def\mbar{{\rm\,M_{baryon}}}
\def\lsun{{\rm\,L_\odot}}
\def\mdot{{\rm\,M_\odot}}
\def\vol#1  {{{#1}{\rm,}\ }}
\def\mytau{\tau_{{\rm Ly}\alpha}}
\def\hi{{H\thinspace I}\ }
\def\hii{{H\thinspace II}\ }
\def\hei{{He\thinspace I}\ }
\def\heii{{He\thinspace II}\ }
\def\heiii{{He\thinspace III}\ }
\def\nhi{N_{HI}}
\def\nheii{N_{\heii}}
\def\jhi{J_{\hi}}
\def\jhiu{J_{-21}}
\def\jheii{J_{\heii}}
\def\shi{\sigma_{\hi}(\nu)}
\def\shia{\bar\sigma_{\hi}}
\def\etal{et al.\ }
\def\eq{eq.\ }
\def\Sec{\S }
\def\eqs{eqs.\ }
\def\cf{{cf.}\ }
 
\def\eqs{eqs.\ }

\def\ni{\noindent}        
\def\ub{\underbar}
\def\hi{\noindent \hangindent=2.5em}
\def\pc{{\rm\,pc}}
\def\kpc{{\rm\,kpc}}
\def\Mpc{{\rm\,Mpc}}
\def\mpc{{\rm\,Mpc}}
\def\hnot{{\rm\,km/s/Mpc}}
\def\kmsec{{\rm\,km/s}}
\def\kms{{\rm\,km/s}}
\def\msun{{\rm\,M_\odot}}
\def\lsun{{\rm\,L_\odot}}
\def\mdot{{\rm\,M_\odot}}
\def\vol#1  {{{#1}{\rm,}\ }}
\def\mytau{\tau_{{\rm Ly}\alpha}}
\def\hi{{H\thinspace I}\ }
\def\hii{{H\thinspace II}\ }
\def\hei{{He\thinspace I}\ }
\def\heii{{He\thinspace II}\ }
\def\heiii{{He\thinspace III}\ }
\def\nhi{N_{HI}}
\def\nheii{N_{\heii}}
\def\jhi{J_{\hi}}
\def\jhiu{J_{-21}}
\def\jheii{J_{\heii}}
\def\shi{\sigma_{\hi}(\nu)}
\def\shia{\bar\sigma_{\hi}}
\def\etal{et al.\ }
\def\eq{eq.\ }
\def\Sec{\S }
\def\eqs{eqs.\ }
\def\cf{{cf.}\ }

\def\ni{\noindent}        
\def\ub{\underbar}
\def\hi{\noindent \hangindent=2.5em}
\def\pc{{\rm\,pc}}
\def\kpc{{\rm\,kpc}}
\def\Mpc{{\rm\,Mpc}}
\def\mpc{{\rm\,Mpc}}
\def\hnot{{\rm\,km/s/Mpc}}
\def\kmsec{{\rm\,km/s}}
\def\cm{\rm cm}
\def\zsun{{\rm\,Z_\odot}}
\def\kms{{\rm\,km/s}}
\def\msun{{\rm\,M_\odot}}
\def\lsun{{\rm\,L_\odot}}
\def\mdot{{\rm\,M_\odot}}
\def\vol#1  {{{#1}{\rm,}\ }}
\def\mytau{\tau_{{\rm Ly}\alpha}}
\def\hi{{H\thinspace I}\ }
\def\hii{{H\thinspace II}\ }
\def\hei{{He\thinspace I}\ }
\def\heii{{He\thinspace II}\ }
\def\heiii{{He\thinspace III}\ }
\def\nhi{N_{HI}}
\def\nheii{N_{\heii}}
\def\jhi{J_{\hi}}
\def\jhiu{J_{-21}}
\def\jheii{J_{\heii}}
\def\shi{\sigma_{\hi}(\nu)}
\def\shia{\bar\sigma_{\hi}}
\def\etal{et al.\ }
\def\eq{eq.\ }
\def\Sec{\S }
\def\eqs{eqs.\ }
\def\cf{{cf.}\ }

\def\clock{\count0=\time \divide\count0 by 60
     \count1=\count0 \multiply\count1 by -60 \advance\count1 by \time
     \number\count0:\ifnum\count1<10{0\number\count1}\else\number\count1\fi}
\def\draft{
   \rightline{DRAFT: \today \quad\quad \clock}}

\def\Kel{\hskip 2pt K\hskip 3pt}
\def\cf{{\it cf.}\hskip 1.5pt}
\def\refset{\parindent=0pt\hangafter=1\hangindent=1em}
\def\refskip{\vskip -0.1cm}

\font\eightrm=cmr8 scaled \magstep0
\def\fig #1, #2, #3, #4, #5, #6 {
\topinsert
\smallskip
\centerline{\psfig{figure=#1,height=#2 in,width=#3 in,angle=#4}}
\medskip
{\vskip #5 cm\leftskip2.5em \parindent=0pt {\eightrm #6 }}
\endinsert}

\def\h2{${\rm\,H_2}$}

\title{\bf CosmoMHD: A Cosmological Magnetohydrodynamics Code}
\author{Shengtai Li\altaffilmark{1}, Hui Li\altaffilmark{2}, 
and Renyue Cen\altaffilmark{3}}
\altaffiltext{1}{
Theoretical Division, MS B284, Los Alamos National
Laboratory, NM 87545; sli@lanl.gov}
\altaffiltext{2}{
Theoretical Division, MS B227, Los Alamos National
Laboratory, NM 87545; hli@lanl.gov}
\altaffiltext{3}{
Department of Astrophysical Sciences, Princeton University,
Peyton Hall - Ivy Lane, Princeton, NJ 08544; cen@atro.princeton.edu}

\begin{abstract}
In this era of precision cosmology, a detailed physical understanding
on the evolution of cosmic baryons is required. Cosmic magnetic
fields, though still poorly understood, may represent an important
component in the global cosmic energy flow that affects the baryon
dynamics. We have developed an Eulerian-based cosmological
magnetohydrodynamics code (CosmoMHD) with modern shock capturing
schemes to study the formation and evolution of cosmic structures in
the presence of magnetic fields.  The code solves the ideal MHD
equations as well as the non-equilibrium rate equations for multiple
species, the Vlasov equation for dynamics of collisionless particles,
the Poisson's equation for the gravitational potential field and the
equation for the evolution of the intergalactic ionizing radiation
field.  In addition, a detailed star formation prescription and
feedback processes are implemented.  Several methods for solving the
MHD by high-resolution schemes with finite-volume and
finite-difference methods are implemented.  The divergence-free
condition of the magnetic fields is preserved at a level of computer
roundoff error via the constraint transport method.  We have also
implemented a high-resolution method via dual-equation formulations to
track the thermal energy accurately in very high Mach number or high
Alfven-Mach number regions. Several numerical tests have demonstrated
the efficacy of the proposed schemes.
\end{abstract}


\section{Introduction}





While the concordance cold dark matter cosmological model
($\Lambda$CDM; Krauss \& Turner 1995; Ostriker \& Steinhardt 1995;
Bahcall \etal 1999; Spergel \etal 2006) has been shown to be
remarkably successful in many respects, there remain several apparent
discrepancies between the model and the real Universe.  For example,
the centers of simulated X-ray clusters of galaxies do not agree with
observations, manifested in their inability to match even the simplest
relations such as the temperature-luminosity relation (e.g., Ponman,
Cannon, \& Navarro 1999), among others.  Closely related to this
problem is the so-called ``cooling flow" problem in the cores of many
clusters, where the cooling time is less than the Hubble time (e.g.,
Fabian 1994; White, Jones, \& Forman 1997; Peres \etal 1998; Allen
2000) and, hence, in the absence of heating sources, the gas will cool
and flow towards the center.  This is confirmed by simple hydrostatic
models (Fabian \& Nulsen 1977; Cowie \& Binney 1977) and simulations
(Suginohara \& Ostriker 1998; Yoshida \etal 2002), with conventional
physics, i.e., gravity and cooling.  What is most puzzling is that,
observationally, this supposedly cooling X-ray gas shows up neither in
soft X-rays nor in other cooler forms in the expected amounts (e.g.,
Peterson et al. 2003; Kaastra et al. 2004).  Solutions such as
supernova heating appear incapable of rectifying the problem (Binney
2000).

Based on a polytropic model taking into account both a removal of low
entropy gas due to star formation and an addition of feedback energy,
Ostriker \etal (2005) show that if the feedback energy is of the order
of a few percent of the rest mass of Supermassive black holes, a
consistent X-ray cluster model can be constructed.  Supermassive black
holes (SMBH) are now strongly believed to lurk at the centers of
probably every massive galaxy (Magorrian \etal 1998; Ferrarese \&
Merritt 2000; Gebhardt \etal 2000; Tremaine \etal 2002).  One of the
most prominent known forms of feedback from the growth of SMBH that is
mechanically tightly coupled to surrounding gas is the powerful radio
jets and giant radio lobes, whose energy is often observed to exceed
$10^{5-6}\msun c^2$ergs (Urry 2000; Kronberg et al. 2001).  Therefore,
there is a good reason to believe that the inclusion of Active
Galactic Nuclei (AGN) feedback, in form of radio jets/lobes, in direct
simulations may bring theoretical predictions into good agreement with
observations.  In this case, our understanding of X-ray cluster
formation will be much more complete and physically sound, which in
turn would allow us to remove systematic uncertainties with regard to
determinations of dark matter and dark energy, among others, afforded
by accumulating X-ray data and upcoming Sunyaev-Zel'dovich (SZ)
cluster data.  Clearly, it has become very pressing to properly model
AGN feedback in cosmological simulations.

In this era of precision cosmology, multiple, independent constraints
on cosmological parameters are vital.  Weak gravitational lensing of
galaxies by large-scale structure provides one of the most powerful
probes of matter distribution in the Universe (e.g., Refregier 2003).
Since weak lensing directly probes the mass distribution, which is
dominated by dark matter, it is often assumed that baryonic physics is
not that important.  With powerful future weak lensing surveys such as
PanSTARRS, SNAP and LSST, which have the potential capability of
reducing statistical errors to about $1\%$ level with regard to power
spectrum and dark energy equation-of-state determinations, it will be
necessary to understand hitherto ``less important" effects such as the
physics of cosmic baryons.  Baryons constitute a significant fraction
($\sim 17\%$) of total mass in the Universe (Spergel \etal 2006) and
could make a significant contribution to the total matter power
spectrum.  On sub-megaparsec scales, baryons are known to be subject
to different physics than the dark matter, and important differences
between distributions of baryons and dark matter on small scales ($\le
100$kpc) are well observed, such as in our own galaxy.  It is,
however, often assumed that baryons follow dark matter on large scales
($\ge 1$Mpc), where signals of weak gravitational lensing by
large-scale structure are detected to in turn determine the total
matter distribution on these large scales.  Estimates based on
observed radio-lobe energy from AGN suggest that a large volume
fraction of the intergalactic medium may be moved and filled with
magnetic bubbles (Furlanetto \& Loeb 2001; Kronberg et al. 2001;
Levine \& Gnedin 2005).  As a result, the effect of large-scale
movement of baryonic gas under the influence of giant radio lobes/jets
from AGN on the total matter power spectrum might be as large as
$30\%$ based on a simplified spherical model (Levine \& Gnedin 2006).
While the exact effect on the intergalactic medium of giant radio
jet/lobes from AGN is presently highly uncertain due to lack of any
adequate treatment, these initial rough estimates of potential effects
point to the need of much more detailed investigations.  It is prudent
to re-examine with greater care the assumption that baryons follow
dark matter on large scales in light of the potential capabilities of
future weak lensing surveys.  Cosmological parameters determined with
claimed achievable high statistical accuracy based on gravitational
weak lensing can be realized only after systematic effects on baryons
due to powerful AGN feedback can be convincingly demonstrated and
removed.  Magnetohydrodynamics simulations provide the necessary tools
to tackle this important problem.

The idea of AGN feedback is not new (e.g., Silk \& Rees 1998; Kronberg
\etal 2001; Churazov \etal 2001; Quilis, Bower, \& Balogh 2001;
Bruggen \& Kaiser 2001; Binney 2004; Omma \& Binney 2004; Ruszkowski,
Bruggen \& Begelman 2004; Henrik \etal 2004; Scannapieco \& Oh 2004;
Begelman 2004; Dalla Vecchia \etal 2004; Begelman \& Ruszkowski 2005;
Bruggen, Ruszkowski, \& Hallman 2005; Croton \etal 2006).  Recent SPH
(smoothed-particle-hydrodynamics) simulations have taken a major step
to implement such an AGN feedback (Hopkins \etal 2005a,b,c,d,2006;
Robertson \etal 2006; Sajacki \& Springel 2006) through a deposition
of thermal energy in the central regions of galaxies where massive gas
accretion onto the central SMBH takes place.  The effects are clearly
demonstrated to be significant, often with dramatic effects, such as a
complete sweep-out of interstellar medium.  But real MHD modeling is
largely lacking.  Undoubtedly radio jets are highly anisotropic and
the overall dynamics and thermodynamics of magnetic jets/bubbles and
baryons on scales ranging from cluster cores to large-scale structure
requires solving the propagation of relativistic jets using
cosmological MHD codes in a complex, dynamic, cosmological setting.
As an example, a highly collimated relativistic radio jet may not
deposit most of its energy in the immediate neighborhood, whereas a
thermal and spherical deposition of the energy, as has been
implemented in SPH simulation, would cause an interaction with the
immediate surrounding gas at the outset.

It has been well established that there are widespread magnetic fields
in the intracluster medium (see Carilli \& Taylor 2002; Govoni 2006
for recent reviews). There has also been tantalizing evidence for
magnetic fields in the wider intergalactic medium (IGM) (Kim et
al. 1989). The origin of these large-scale magnetic fields is still
unknown though several suggestions have been made (e.g., Kulsrud et
al. 1997; Furlanetto \& Loeb 2001; Kronberg et al. 2001). Furthermore,
it is presently mostly unknown whether these magnetic fields have
played an important role or not, in systems ranging from large-scale
structure formation, to galaxy clusters, to cluster core regions and
to galaxy formation itself. There have been some studies aimed at
investigating the role of magnetic fields in the intracluster (ICM)
and IGM (Ryu et al. 1998; Dolag et al. 2002; Dolag et al. 2005;
Br\"uggen et al. 2005). Numerically, substantial efforts have gone
into developing MHD solvers with SPH (Dolag et al. 1999, 2002), though
the divergence-free condition for magnetic fields is still difficult
to handle (Ziegler et al. 2006). For a grid-based Eulerian approach,
studies by Ryu et al. (1998) and Br\"uggen et al. (2005) have included
the magnetic fields passively, i.e., they have omitted the feedback of
magnetic fields to the medium (both in the momentum equation and
energy equation). Consequently, we are not aware of any Eulerian-based
cosmological simulations where magnetic fields are included
self-consistently.

In this paper, we present a highly accurate cosmological MHD code,
named CosmoMHD, for cosmological simulations that involve magnetic
fields.  The outline of our paper is as follows. We first present an
overview of the CosmoMHD code and various physics packages included in
CosmoMHD in \S \ref{sec:eqn}. We then describe the basic solvers for
ideal MHD adopted in our code in \S \ref{sec:solver}. Numerical tests
are presented in \S \ref{sec:tests}. Further discussions are given in
\S \ref{sec:conclu}.

\section{\label{sec:eqn}CosmoMHD Code and Physics Packages}        

\subsection{CosmoMHD Code}

The framework of our CosmoMHD code is based on the combination of the
TVD-ES cosmological hydrodynamics (HD) code (Ryu \etal 1993) and an MHD
code on an adaptive mesh refinement (AMR-MHD) grid (Li \& Li 2003).
We have integrated the cosmological solvers for dark matter, atomic
physics and star formation feedback processes in the TVD-ES code with
the MHD solvers in AMR-MHD code for gas dynamics (AMR features are
{\it not} utilized in CosmoMHD).  The hydro TVD-ES code has been
extensively used for cosmological applications, including Lyman-alpha
forest (Cen \etal 1994), warm-hot intergalactic medium (Cen \&
Ostriker 1999a), damped Lyman-alpha systems (Cen \etal 2003),
cosmological chemical evolution (Cen \& Ostriker 1999b; Cen, Nagamine,
\& Ostriker 2005) and galaxy formation (Nagamine \etal
2000,2001a,b,2004,2005a,b,2006).  The MHD code has been extensively
used for astrophysical jet simulations (e.g., Li et al. 2006,
Nakamura, Li, \& Li 2006).

Consistent with the notations in Ryu et al. (1993), the basic
equations for CosmoMHD in comoving coordinates are
\begin{eqnarray}
\frac{\partial\rho}{\partial t} + \frac{1}{a}\nabla\cdot(\rho\mbox{\bf
v}) &=& 0 \label{cmhd_eq1}\\
\frac{\partial (\rho {\bf v})}{\partial t} + \frac{1}{a}\nabla\cdot
\left(\rho {\bf v \bf v} + p^* - {\bf B}{\bf B}\right) &=& 
- \frac{\dot a}{a} \rho {\bf
v} - \frac{1}{a} \rho \nabla \Phi \\
\frac{\partial E}{\partial t} + \frac{1}{a}\nabla\cdot[(E +
p^*)\mbox{\bf v} - ({\bf B}\cdot{\bf v}){\bf B}] & = & 
- \frac{2\dot a}{a} E  - \frac{1}{a} \rho {\bf v} \cdot \nabla \Phi +
\frac{\dot a}{a} \frac{B^2}{2}, \\
\frac{\partial {\bf B}}{\partial t} - \frac{1}{a}\nabla\times(\mbox{\bf
v}\times {\bf B}) &=& - \frac{\dot a}{2a} {\bf B} \label{cmhd_eq4}
\end{eqnarray}
where we have implicitly assumed $\gamma = 5/3$. Here, $\rho$ is the
comoving density, ${\bf v}$ is the proper peculiar velocity, $p$ is
the comoving gas pressure, $B^2/2$ is the comoving magnetic pressure,
$p^* = p + B^2/2$ is the total comoving pressure, $E = \frac12\rho
\mbox{v}^2 + \frac{3}{2} p+\frac12 B^2$ is the total peculiar energy
per unit comoving volume, $\Phi$ is the proper peculiar gravitational
potential from both dark matter and self-gravity, $t$ is the cosmic
time, and $a$ is the expansion parameter.  Note that, from the
induction equation above [Eq. (\ref{cmhd_eq4})], the magnetic field in
the comoving frame expands as $B_{\rm comoving} \propto
a^{-1/2}$. Since $B_{\rm comoving} = B_{\rm proper} a^{3/2}$, the
magnetic field in the proper frame expands as $B_{\rm proper} \propto
a^{-2}$, consistent with the flux conservation.

In addition to the total energy equation, we have implemented solvers
for two auxiliary equations: the modified entropy equation given in
Ryu et al. (1993) and the internal energy equation in Bryan et
al. (1995). This implementation allows us to perform better
comparisons among different approaches in tracking the thermal energy
accurately. The equations are:
\begin{eqnarray}
\frac{\partial S}{\partial t} + \frac{1}{a}\nabla\cdot(S{\bf v}) 
&=& - \frac{2\dot a}{a} S~, \label{mentropy}\\
\frac{\partial\rho e}{\partial t} + \frac{1}{a}\nabla\cdot(\rho e{\bf
v}) &=& -\frac{2\dot a}{a} \rho e + \frac{p}{a}\nabla\cdot{\bf v},
\label{eq_ein_den} 
\end{eqnarray}
where $S\equiv p/\rho^{\gamma-1}$ is the comoving modified entropy,
and $e$ is the internal energy.  Note that Eqs. (\ref{mentropy}) and
(\ref{eq_ein_den}) remain the same even when the magnetic fields are
present.


The CosmoMHD code integrates five sets of equations simultaneously:
the ideal MHD equations for gas dynamics and magnetic field dynamics
[Equations (\ref{cmhd_eq1}) to (\ref{eq_ein_den})], rate equations for
multiple species of different ionizational states (including hydrogen,
helium and oxygen), the Vlasov equation for dynamics of collisionless
particles, the Poisson's equation for obtaining the gravitational
potential field and the equation governing the evolution of the
intergalactic ionizing radiation field, all in cosmological comoving
coordinates.  The MHD code consists of several approaches for solving
the MHD (or HD) by high-resolution schemes with finite-volume and
finite-difference methods.  The code preserves conservative quantities
and the divergence-free condition of the magnetic fields.  The code is
fully parallelized with OPENMP directives.  The MHD solver can also be
used as an hydro solver when the magnetic fields are zero. We will
describe the MHD solvers in more details in \S \ref{sec:solver}.  The
rate equations are treated using sub-cycles within a hydrodynamic
timestep due to much shorter ionization timescales.  Dark matter
particles are advanced in time using the standard particle-mesh (PM)
scheme.  The gravitational potential on a uniform mesh is solved
using the Fast Fourier Transform (FFT) method.

The simulation computes for each cell and each timestep the detailed
cooling and heating processes due to all the principal line and
continuum processes for a plasma of primordial composition (Cen 1992).
Metals ejected from star formation are followed in detail in a
time-dependent, non-equilibrium fashion.  Cooling due to metals is
computed using a code based on the Raymond-Smith code assuming
ionization equilibrium (Cen \etal 1995): at each timestep, given the
ionizing background radiation field, we compute a lookup table for
metal cooling in the temperature-density plane for a gas with solar
metallicity, then the metal cooling rate for each gas cell is computed
using the appropriate entry in that plane multiplied by its
metallicity (in solar units).  In addition, Compton cooling due to the
microwave background radiation field and Compton cooling/heating due
to the X-ray and high energy background are also included.

We follow star formation using a well defined (heuristic but plausible)
prescription used by us in our earlier work
(Cen \& Ostriker 1992,1993) and 
similar to that of other investigators 
(Katz, Hernquist, \& Weinberg 1992;
Katz, Weinberg, \& Hernquist 1996;
Steinmetz 1996;
Gnedin \& Ostriker 1997).
A stellar particle of mass
$m_{*}=c_{*} m_{\rm gas} \Delta t/t_{*}$ is created
(the same amount is removed from the gas mass in the cell),
if the gas in a cell at any time meets
the following three conditions simultaneously:
(i) flow contracting, (ii) cooling time less than dynamic time, and 
(iii) Jeans unstable,
where $\Delta t$ is the timestep, $t_{*}={\rm max}(t_{\rm dyn}, 10^7$yrs),
$t_{dyn}$ is the dynamical time of the cell,
$m_{\rm gas}$ is the baryonic gas mass in the cell and
$c_*\sim 0.07$ is star formation efficiency.
In essence, we follow 
the classic work of Eggen, Lynden-Bell \& Sandage (1962)
and assume that the dynamical free-fall and galaxy formation timescales
are simply related.
Each stellar particle has a number of other attributes at birth, including 
formation time $t_i$, initial gas metallicity
and the free-fall time in the birth cell $t_{dyn}$.
The typical mass of a stellar particle in the simulation
is about one million solar masses;
in other words, these stellar particles are like globular clusters.

Stellar particles are subsequently treated dynamically as
collisionless particles.  But feedback from star formation is allowed
in three forms: ionizing UV photons, supernova kinetic energy in the
form of galactic superwinds (GSW), and metal-enriched gas, all being
proportional to the local star formation rate.  The temporal release
of all three feedback components at time $t$ has the same form:
$f(t,t_i,t_{dyn}) \equiv (1/ t_{dyn})
[(t-t_i)/t_{dyn}]\exp[-(t-t_i)/t_{dyn}]$. Within a timestep $dt$, the
released GSW energy to the IGM, ejected mass from stars into the IGM
and escape UV radiation energy are $e_{GSW} f(t,t_i,t_{dyn}) m_* c^2
dt$, $e_{mass} f(t,t_i,t_{dyn}) m_* dt$ and $f_{esc}(Z) e_{UV}(Z)
f(t,t_i,t_{dyn}) m_* c^2 dt$.  We use the Bruzual-Charlot population
synthesis code (Bruzual \& Charlot 1993; Bruzual 2000) to compute the
intrinsic metallicity-dependent UV spectra from stars with Salpeter
IMF (with a lower and upper mass cutoff of $0.1\msun$ and $125\msun$).
Note that $e_{UV}$ is no longer just a simple coefficient but a
function of metallicity.  The Bruzual-Charlot code gives
$e_{UV}=(1.2\times 10^{-4}, 9.7\times 10^{-5}, 8.2\times 10^{-5},
7.0\times 10^{-5}, 5.6\times 10^{-5}, 3.9\times 10^{-5} ,1.6\times
10^{-6})$ at $Z/\zsun=(5.0\times 10^{-3}, 2.0\times 10^{-2}, 2.0\times
10^{-1}, 4.0\times 10^{-1}, 1.0, 2.5 ,5.0)$.  We also implement a gas
metallicity-dependent ionizing photon escape fraction from galaxies in
the sense that higher metallicity hence higher dust content galaxies
are assumed to allow a lower escape fraction; we adopt the escape
fractions of $f_{esc}=2\%$ and $5\%$ (Hurwitz \etal 1997; Deharveng
\etal 2001; Heckman \etal 2001) for solar and one tenth of solar
metallicity, respectively, and interpolate/extrapolate using a linear
log form of metallicity. In addition, we include the emission from
quasars using the spectral form observationally derived by Sazonov,
Ostriker, \& Sunyaev (2004), with a radiative efficiency in terms of
stellar mass of $e_{QSO}=2.5\times 10^{-5}$ for $h\nu>13.6$eV.
Finally, hot, shocked regions (like clusters of galaxies) which emit
ionizing photons due to bremsstrahlung radiation are also included.
The UV component is simply averaged over the box, since the light
propagation time across our box is small compared to the timesteps.
The radiation field (from $1$eV to $100$keV) is followed in detail
with allowance for self-consistently produced radiation sources and
sinks in the simulation box and for cosmological effects, i.e.,
radiation transfer for the mean field $J_\nu$ is computed with
stellar, quasar and bremsstrahlung sources and sinks due to \lya
clouds etc. In addition, a local optical depth approximation is
adopted to crudely mimic the local shielding effects: each cubic cell
is flagged with six hydrogen ``optical depths" on the six faces, each
equal to the product of neutral hydrogen density, hydrogen ionization
cross section and scale height, and the appropriate mean from the six
values is then calculated; equivalent ones for neutral helium and
singly-ionized helium are also computed. In computing the global sink
terms for the radiation field, the contribution of each cell is
subject to the shielding due to its own ``optical depth".  In
addition, in computing the local ionization and cooling/heating
balance for each cell, we take into account the same shielding 
to attenuate the external ionizing radiation field.

Galactic superwinds energy and ejected metals are distributed into 27
local gas cells centered at the stellar particle in question, weighted
by the specific volume of each cell.  We fix $e_{mass}=0.25$.  GSW
energy injected into the IGM is included with an adjustable efficiency
(in terms of rest-mass energy of total formed stars) of $e_{GSW}$,
which is normalized to observations for our fiducial simulation with
$e_{GSW}=3\times 10^{-6}$.  If the ejected mass and associated energy
propagate into a vacuum, the resulting velocity of the ejecta would be
$(e_{GSW}/e_{mass})^{1/2}c=1469$km/s.  After the ejecta has
accumulated an amount of mass comparable to its initial mass, the
velocity may slow down to a few hundred km/s.  We assume this velocity
would roughly correspond to the observed outflow velocities of Lyman
break galaxies (LBGs) (e.g., Pettini \etal 2002).

We have also implemented a prescription to insert the initial MHD
radio jets associated with the SMBH formation, which in turn is based
on the assumption that SMBH formation is directly related to merger
events.  Details will be presented elsewhere.

\section{\label{sec:solver}Numerical Scheme for Ideal MHD}






\subsection{Ideal MHD Equations}


Without expansion factor and external sources, the Equations 
(\ref{cmhd_eq1}) to (\ref{cmhd_eq4}) can be written in the  
conservative form as  (in Cartesian coordinates):
\begin{equation}
\frac{\partial\mbox{\bf q}}{\partial t} + 
\frac{\partial F}{\partial x_1} + \frac{\partial G}{\partial x_2} +
  \frac{\partial H}{\partial x_3} = 0,
\end{equation}
where $
\mbox{\bf q} = (\rho, \rho v_1, \rho v_2, \rho v_3, B_1, B_2, B_3, E)^t,
$ and the flux functions are
\begin{equation}
\left(\matrix{F\cr G\cr H}\right)^T = \left(\matrix{\rho v_1 & \rho
    v_2 & \rho v_3 \cr 
\rho v_1^2 - B_1^2 + p^* & \rho v_2v_1 - B_2B_1
    & \rho v_3v_1 - B_3B_1 \cr
 \rho v_1v_2 - B_1B_2 & \rho v_2^2 - B_2^2  + p^* & \rho v_3v_2 - B_3B_2 \cr 
\rho v_1v_3 - B_1B_3 & \rho v_2v_3 - B_2B_3 & \rho v_3^2 - B_3^2  +
p^* \cr
 0 & - \Omega_3  & \Omega_2 \cr 
\Omega_3 & 0 & \Omega_1 \cr 
-\Omega_2 & -\Omega_1 & 0\cr
 (E+p^*)v_1 - B_1({\bf B}\cdot\mbox{\bf v}) & (E+p^*)v_2 - B_2({\bf
   B}\cdot\mbox{\bf v}) & (E+p^*)v_3 - B_3({\bf B}\cdot\mbox{\bf v})
}\right).
\end{equation}
All variables carry their usual meaning,
\begin{equation}
p^* = p + \frac12B^2
\end{equation}
is the total pressure, 
\begin{equation}
\Omega_1 = v_2B_3 - v_3B_2, \quad \Omega_2 = v_3B_1 - v_1B_3, \quad
\Omega_3 = v_1B_2 - v_2B_1
\end{equation}
are the electromotive force (EMF, defined via ${\bf \Omega} =
\mbox{\bf v}\times {\bf B}$). We also assume the ideal gas law for equation
of state, and hence the total energy is
\begin{equation}
E = \frac12\rho \mbox{v}^2 + \frac{p}{\gamma-1}+\frac12B^2 ~~~.
\end{equation}

From the flux functions, we can obtain the Jacobian matrices
\begin{equation}
A_x(\mbox{\bf q}) = \frac{\partial F}{\partial \mbox{\bf q}}, \quad
A_y(\mbox{\bf q}) = \frac{\partial G}{\partial \mbox{\bf q}}, \quad
A_z(\mbox{\bf q}) = \frac{\partial H}{\partial \mbox{\bf q}} ~~~. 
\end{equation}

Since $G$ and $H$ can be related to $F$ through proper index
permutation, $A_y$ and $A_z$ should have similar structure as $A_x$.
The eigenvalues and eigenvectors for $A_x$ have been extensively
studied by many authors (see Brio \& Wu 1988, Ryu \& Jones 1995,
etc.). Here we will write out the eigenvalues without explanation. In
a direct extension of the 1D system with magnetic field component
$B_1$ and total magnetic field ${\bf B}$, the eigenvalues for
$A_x(\mbox{\bf q})$ are
\begin{equation}
\lambda_{1,7} = v_1 \pm v_f, \quad \lambda_{2,6} = v_1 \pm v_a, \quad
\lambda_{3,5} = v_1 \pm v_s, \quad \lambda_4 = v_1,
\end{equation}
where $ v_a = \sqrt{{B_1^2}/{\rho}} $ 
is the Alfv\'en speed based on $B_1$, and 
\begin{equation}
v_f = \left[\frac12\left(c_s^2 + \frac{{\bf B}^2}{\rho} +
    \sqrt{\left(c_s^2+\frac{{\bf B}^2}{\rho}\right)^2 -
      4c_s^2v_a^2}\right)\right]^\frac12,  
\end{equation}
\begin{equation}
v_s = \left[\frac12\left(c_s^2 + \frac{{\bf B}^2}{\rho} -
    \sqrt{\left(c_s^2+\frac{{\bf B}^2}{\rho}\right)^2 - 
    4c_s^2v_a^2}\right)\right]^\frac12, 
\end{equation}
are the speeds of fast and slow magneto-sonic waves respectively, and $c_s=
\sqrt{{\gamma p}/{\rho}}$ is the  
acoustic wave speed.

In an alternative formulation, Powell et al. (1999) have used an
$8\times 8$ eigen-system. Its corresponding eigenvalues are 
\begin{equation}
\lambda_{1,8} = v_1 \pm v_f, \quad \lambda_{2,7} = v_1 \pm v_a, \quad
\lambda_{3,6} = v_1 \pm v_s, \quad \lambda_{4,5} = v_1 ~~~.
\end{equation}
The corresponding eigenvectors have been given previously 
by many authors. We have adopted the eigenvector set given by Powell et
al. (1999) in our code. 

For the dimensional split approach, the CFL condition along each
direction can be calculated with (assume the same timestep is used)
\begin{equation}
\mbox{CFL} = dt\cdot\min\left[\frac{|v_1| + v_f}{dx_1}, 
\frac{|v_2| + v_f}{dx_2}, \frac{|v_3| + v_f}{dx_3}\right], 
\end{equation}
where the minimum is over all the cells. For the unsplit approach, the
CFL condition is given as
\begin{equation}
\mbox{CFL} = dt\cdot\min\left[\frac{|v_1| + v_f}{dx_1}+
\frac{|v_2| + v_f}{dx_2}+ \frac{|v_3| + v_f}{dx_3}\right], 
\end{equation}
where again, the minimum is over all the cells.

\subsection{Numerical Solvers for Ideal MHD}

In recent years a variety of numerical algorithms for MHD based on the
Godunov method have been developed. Early development of higher order
Godunov schemes for MHD focused on interpreting the MHD equations as a
simple system of conservation laws. This was done by Brio and Wu
(1988), Zachary, Malagoli and Colella (1994), Powell (1994), Dai and
Woodward (1994), Ryu and Jones (1995), Roe and Balsara (1996),
etc. There are two important extensions to the basic HD algorithm for
use of MHD. The first is an extension of the Riemann solver used to
compute the fluxes of each conserved quantity to MHD; the second is to
preserve the divergence-free constraints $\nabla\cdot {\bf B} = 0$
numerically.

A typical second-order Godunov scheme for hyperbolic conservation
law consists of three steps:
\begin{enumerate}
\item Predict the time-centered cell-interface values, $U^{n+1/2}_L$ and
  $U^{n+1/2}_R$.
\item Using these Left and Right states at each cell-interface, solve the
  Riemann problem to determine the time-centered fluxes,
  $F(U^{n+1/2}_L,U^{n+1/2}_R)$, possibly modified by the additional
  artificial viscosity.
\item Compute the cell-centered conservative variable at the next time
  step using a conservative update with the time-centered fluxes.
\end{enumerate}
The first step is a predictor step via data reconstruction and time
evolution. The second step is a Riemann solver.

\subsubsection{\label{sec:recs}Data Reconstruction}

We included several data reconstruction approaches in our code.
We used piecewise linear reconstruction along with several slope
limiters, including a general {\it minmod}, 
\begin{equation}
\overline{\Delta u_i} = \mbox{minmod}\left(\theta( u_{i+1}-u_{i}),
\frac12(u_{i+1}-u_{i-1}), \theta(u_i - u_{i-1})\right),
\end{equation}
where $\overline{\Delta u_i}$ is a limited slope for cell
$[x_{i-\frac12},x_{i+\frac12}]$, and $\theta\in[1,2]$. Note that a
larger $\theta$ corresponds to less dissipation, but still
non-oscillatory limiters. For $\theta = 2$, it becomes a Woodward
limiter. These minmod-type limiters are not smooth functions with
respect to $u$.  Another limiter we used is van Albada-type, which is
a smooth function of $u$,
\begin{equation}
\overline{\Delta u_i} = \frac{2(u_{i+1}-u_i)(u_i-u_{i-1}) + \varepsilon}
  {(u_{i+1}-u_i)^2 + (u_i - u_{i-1})^2 + \varepsilon}\frac12(u_{i+1}-u_{i-1}),
\end{equation}
where $\varepsilon$ is a tiny positive constant in case of
$u_i=u_{i+1}$ and $u_i=u_{i-1}$. 
This limiter is smooth and it preserves the monotonicity of the
original profile of $u$. 

For robustness, we have also implemented the characteristic limiting
data reconstruction, which applies the limited slope on wave-by-wave
decomposition space obtained from local solutions of Riemann
problems. 
In addition, we have also implemented the piecewise parabolic (Colella
\& Woodward 1984) and central WENO (Levy et al. 2000; 2002)
reconstruction schemes. For robustness, the
data reconstruction is applied to the primitive variables
rather than the conservative variables.

\subsubsection{Predictor for Time-centered Quantities}

We have implemented several approaches as the predictor for the
time-centered quantities. The simplest approach is the Hancock
predictor,
\begin{eqnarray*}
u_{i}^{n+\frac12} = u^n_{i} &+& \frac12\frac{\Delta t}{\Delta
  x}\left(F(u_{i}^n + \frac12\overline{\Delta_x u_{i}^n}) - F(u_{i}^n
- \frac12\overline{\Delta_x u_{i}^n})\right).
\end{eqnarray*}
where $\overline{\Delta_{x}u^n_{i}}$ is the limited slope
that is obtained from the data reconstruction. Then we use
$u_{i}^{n+\frac12}$ and 
$\overline{\Delta_{x}u^n_{i}}$ to construct the left and right
interface values at time $t^{n+\frac12}$. In $x$ direction, we have
\begin{equation}
u_{i+\frac12}^L = u_{i}^{n+\frac12} +
\frac12\overline{\Delta_{x}u^n_{i}}, \quad 
u_{i+\frac12}^R = u_{i+1}^{n+\frac12} -
\frac12\overline{\Delta_{x}u^n_{i+1}}. 
\end{equation}
Since the Hancock prediction does not need the characteristic
decomposition, it can be done fast. Note that the limited
slopes are reconstructed only once during the whole
process. We do not need to calculate the limited slope for
intermediate state $u^{n+1/2}$. 

We have also implemented the piecewise-linear method (PLM) via
upwinding characteristic tracing (Colella and Glaz, 1985).

\subsubsection{\label{sec:riem}Riemann solvers}

We implemented three different Riemann solvers, including Roe (Powell
et al. 1999), HLL/HLLE [see Harten, Lax, \& van Leer (1983); Einfeldt
(1988); Einfeldt et al. (1991)], and HLLC (Li 2005). To improve the
robustness, we also combined the HLLE and Roe's Riemann solves
together in our code as proposed by Janhunen (2000). 

\subsection{Dual Energy Formulation for High Mach and Low-$\beta$
  Plasma Flows}

In cosmological simulations, hypersonic flows with a large Mach number
(e.g., $M\geq 100$) appear to be very common, and they present a
problem in our numerical schemes so far because the thermal energy is
very small compared to the kinetic energy.  The pressure, which is
proportional to the thermal energy and used extensively in the
numerical schemes, cannot be tracked accurately, because the
discretization errors made in computing the total energy and the
kinetic energy can be large enough to result in negative pressure. A
similar problem arises in modeling the MHD flow with a low plasma beta
($\beta =2P/{\bf B}^2$).

Dynamically, this is not a large problem because the negative pressure
can be replaced with a nominal floor value and the total energy budget
of the flow remains unaffected. However, if the temperature
distribution is considered for other reasons, (e.g., radiative
processes), the thermal energy must be tracked accurately. 

As suggested by Ryu et al. (1993) and Bryan et al. (1995), we present dual 
energy formulations to track the thermal energy accurately. For comparison, 
we implement two approaches. The 
first is to solve the modified entropy Eq. (\ref{mentropy}), which is 
in conservative form and easy to implement for solvers that require 
eigen-decomposition. The other is to solve the internal energy density 
equation in Eq. (\ref{eq_ein_den}).

A similar set of Riemann solvers described in \S (\ref{sec:riem})
are also implemented when solving the dual energy formulation. Since
the internal energy equation is not in conservative form, we
calculated only the flux $\nabla\cdot(\rho e{\bf v})$ in
Eq. (\ref{eq_ein_den}) with the Riemann solver.  The Roe's solver
requires eigen-decomposition for a conservative system and cannot be
applied to Eq. (\ref{eq_ein_den}) directly.  For the modified entropy
equation (\ref{mentropy}), we adopt the eigen-decomposition from
Balsara and Spicer (1999b) in our Roe's solver. The HLL and HLLC
solvers can be applied to flux $\nabla\cdot(\rho e{\bf v})$ as well as
to Eq. (\ref{mentropy}) in a straightforward manner. The $p\nabla\cdot
{\bf v}$ in Eq. (\ref{eq_ein_den}) can be discretized as
\begin{equation}
(p\nabla\cdot {\bf v})_{i}^{n+1/2} = \frac{p_{i+1/2}^{n+1/2} +
  p_{i-1/2}^{n+1/2}}{2}(v_{i+1/2}^{n+1/2}-v_{i-1/2}^{n+1/2})/\Delta x
\end{equation}
where the time-centered values of $p_{i+1/2}^{n+1/2}$ and $v^{n+1/2}_{i+1/2}$ 
are
obtained from the Riemann solver. 

It is necessary to use the total energy equation as much as possible
for proper conversion of kinetic to thermal energy. When the total
energy equation fails to give the positive pressure, we switch to
solve the modified entropy or internal energy equation. This is done
by monitoring the values of the internal energy $e$ that comes from
the total energy equation. When the ratio between the internal energy
$e$ and the total energy $E$ satisfies $e/E < \eta$,
Eq. (\ref{mentropy}) or Eq. (\ref{eq_ein_den}) is solved and the total
energy is updated with the new internal energy. We remark that as long
as the parameter $\eta$ is small enough, the dual formulation will
have no dynamical effect. After testing a large number of problems, we
have chosen $\eta=0.008$. In Ryu et al. (1993), $\eta=0.02$ combined
with a shock detection is used. In Bryan et al. (1995), $\eta=0.001$
combined with a shock detection is used. In our code, we tested
$\eta=0.008$ with and without shock detection. Although both
Eqs. (\ref{mentropy}) and (\ref{eq_ein_den}) will not give a correct
solution for the shock, we have not found any problem fails using $e/E
< \eta$ only without shock detection. On the other hand, if $e/E <
\eta$ is used with a shock detection scheme, the internal energy of
the cell near the shock usually is set to a nominal floor value, which
is incorrect.

\subsection{Constraint Transport (CT) for Magnetic Fields}

Maintaining the divergence-free condition $\nabla\cdot{\bf B} =0$ is
important in MHD simulations. Brackbill and Barnes (1980) find that
the non-zero divergence, caused by discretization errors, can grow
exponentially during the computation and destroy the correctness of
the solutions. Hence they proposed a divergence-cleaning approach,
which solves an extra global elliptic (Poisson) equation to recover
$\nabla\cdot {\bf B} = 0$ at each time-step. Balsara \& Kim (2004)
find that the divergence-cleaning is significantly inadequate when
used for many astrophysical applications. The constrained
transport (CT) approach uses a staggered grid, which places the
magnetic field variables at the cell-face to keep $\nabla\cdot {\bf
B}$ to the accuracy of machine round-off error.  Recently, this
approach has been combined with various modern shock-capturing
algorithms by many authors (see Toth 2000 and its references).  Toth
(2000) proposed a constrained transport/central-difference (CT/CD)
method that uses cell-centered values of the magnetic fields. This
method works well with single grid but it is unknown currently how to
apply it to adaptive grids.  In the current stage of CosmoMHD, we deal
with only a single grid. Therefore, we adopted the flux-CT method of
Toth (2000) to preserve the divergence-free condition.

\section{\label{sec:tests}Numerical Tests}
 
\subsection{Shock-tube Test}

We first tested the code with an MHD shock tube problem on a
two-dimensional (2D) Cartesian grid. The shock tube is usually posed
as a 1D problem, and it is a standard test for a numerical scheme to
handle the discontinuities, e.g., shock and contact. Solving it on a
2D grid can verify if a multi-dimensional scheme works properly and
recovers the 1D solution profile.

We chose an example that has been used by Toth (2000) to compare
several numerical schemes for MHD. We adopted the same initial and
boundary conditions as Toth (2000). The initial left and right states
are
$$
(\rho,v_{\|},v_{\perp},v_z,p,B_{\|},B_{\perp},B_z) = \cases{
(1.08,1.2,0.01,0.5, 0.95,
2/\sqrt{4\pi},3.6/\sqrt{4\pi},2/\sqrt{4\pi}), & left, \cr 
(1,0,0,0,1, 2/\sqrt{4\pi},4/\sqrt{4\pi},2/\sqrt{4\pi}), & right. \cr}
$$ where $\parallel$ refers to the direction along the normal of the
shock front, $\perp$ refers to the direction perpendicular to the
normal of the shock front but still in the computational plane, and
$z$ refers to the direction out of the plane.  Note that $v_z$ and
$B_z$ components are non-zero. Therefore, it is often called the 2.5D
MHD shock tube problem. We first solve this problem using a 1D grid
and then solve it as a 2D problem with a rotation angle $\alpha =
45^{\circ}$ between the shock interface and $y$-axes. The same
boundary conditions as Toth (2000) and 256 cells in each direction are
used. The computation is stopped at $t=0.1/\sqrt{2}$ before the fast
shocks reach the left and right boundaries. The numerical solution is
compared with a reference solution that is calculated with a 1D grid
and 1600 cells.

\begin{figure}[htbp]
\begin{center}
\includegraphics[width=0.45\textwidth,height=0.3\textheight]{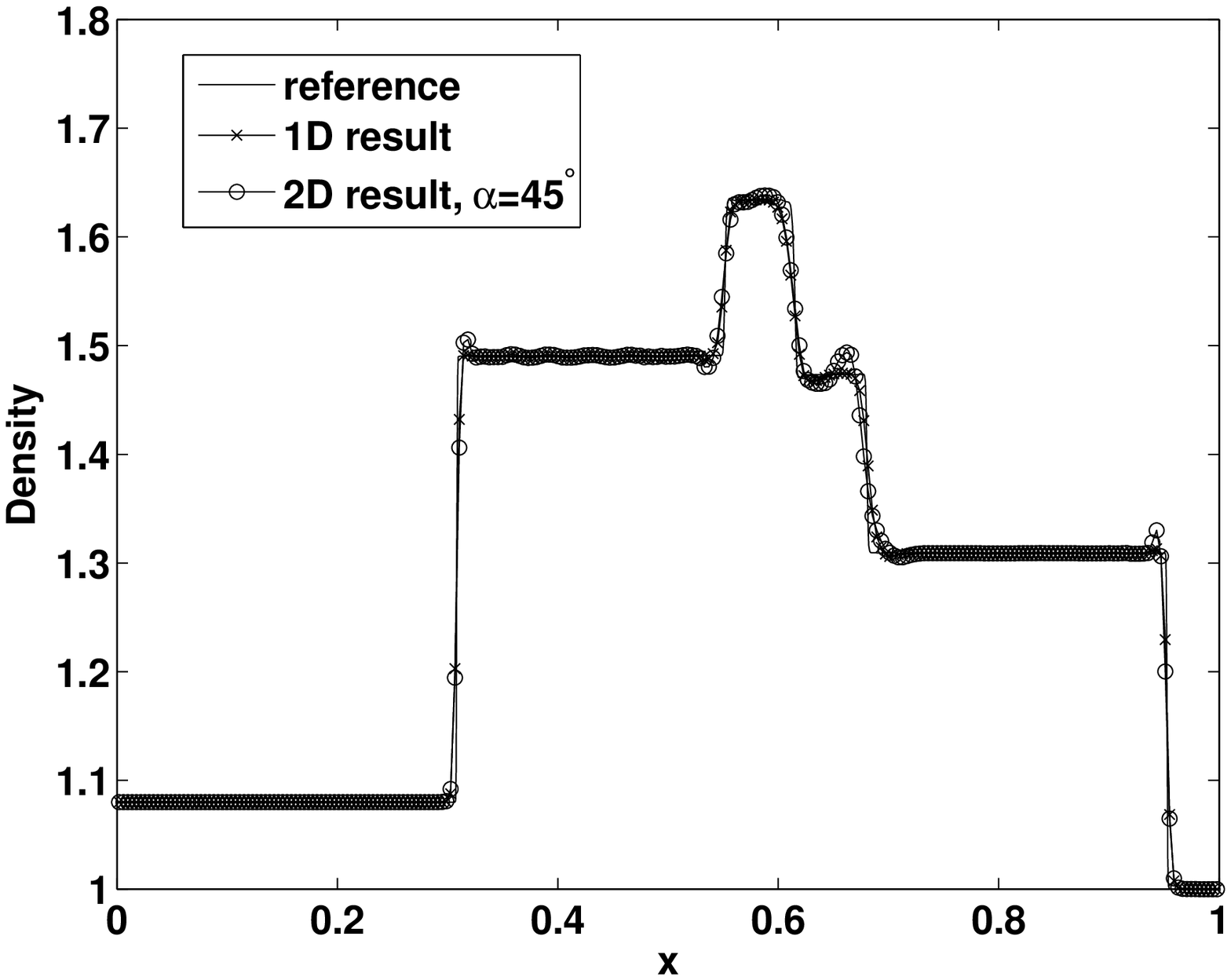}
\includegraphics[width=0.45\textwidth,height=0.3\textheight]{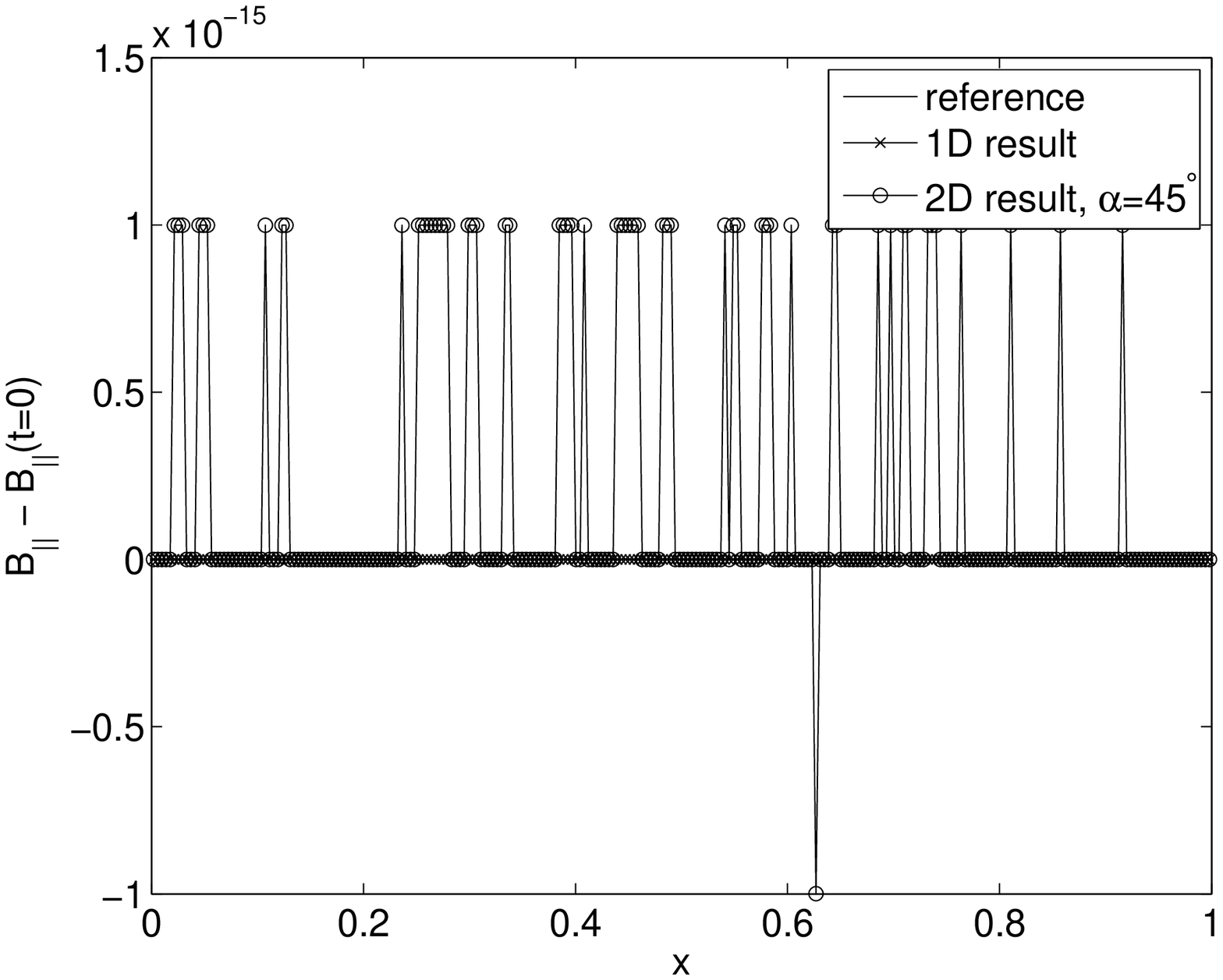}
\includegraphics[width=0.45\textwidth,height=0.3\textheight]{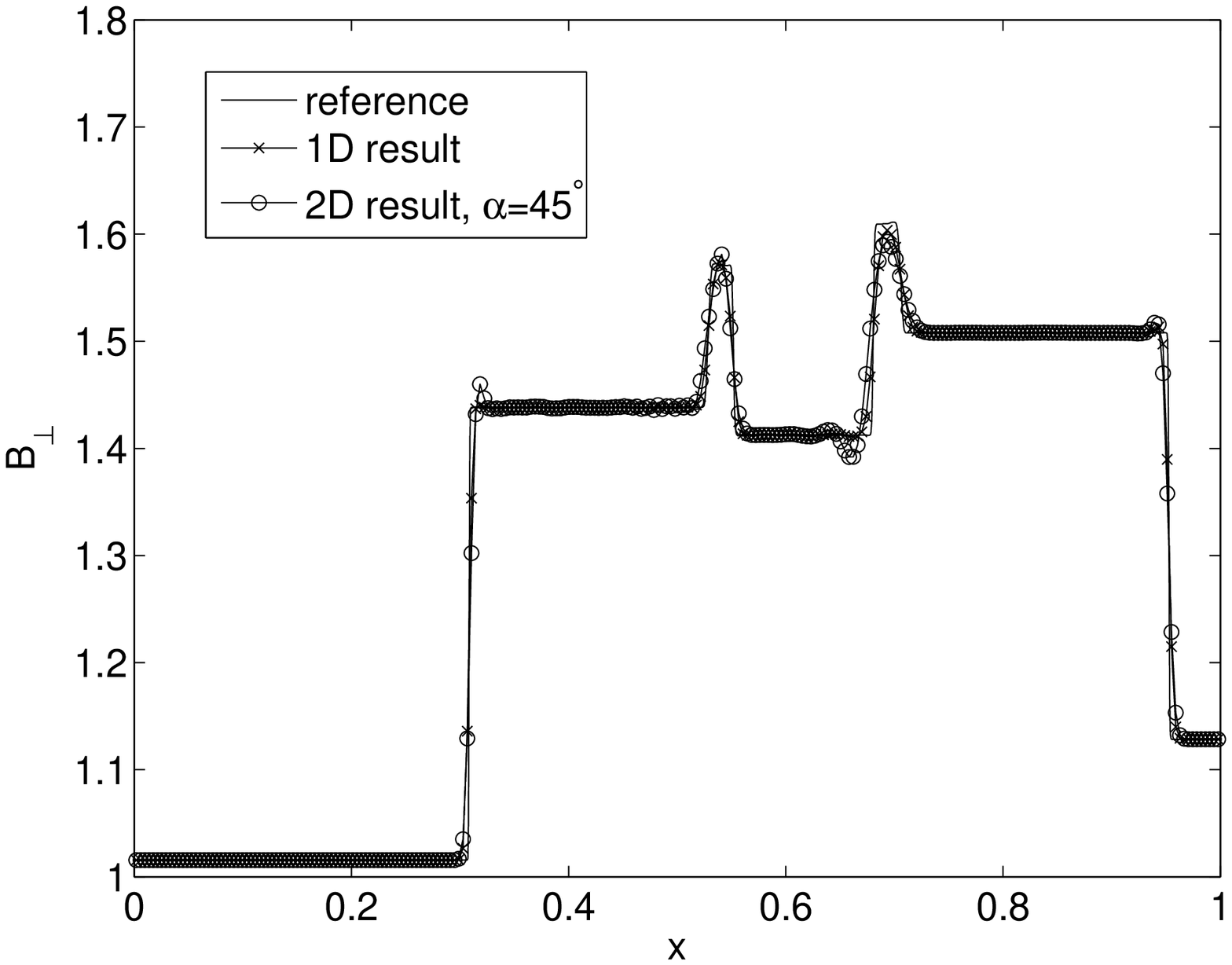} 
\includegraphics[width=0.45\textwidth,height=0.3\textheight]{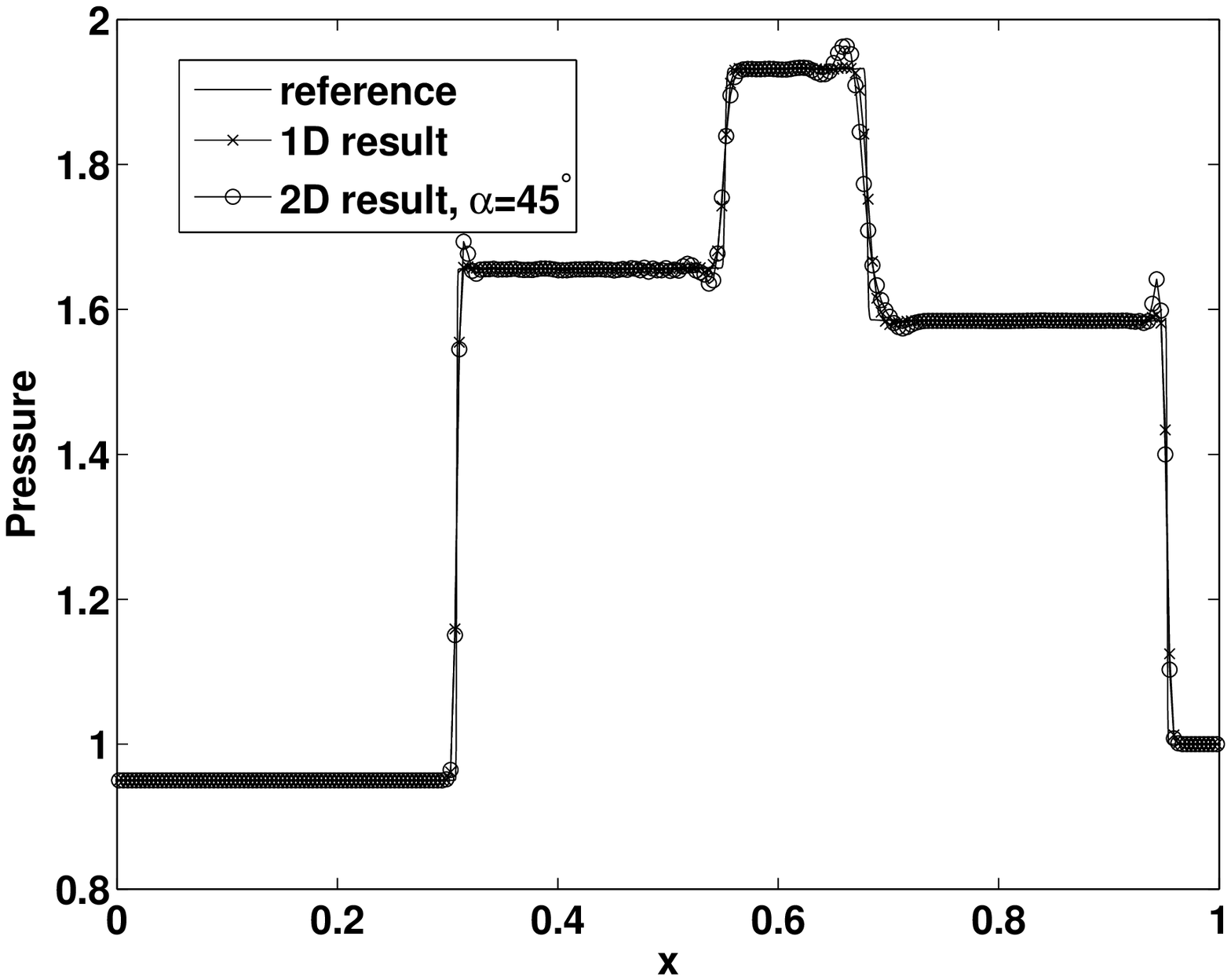} 
\caption{\label{fig1}Plots of density, magnetic components in parallel
and vertical direction along the normal of the shock front, and
pressure of the 2.5D shock-tube problem. The reference solution is
calculated with 1D grid and 1600 grid cells. The 1D and 2D results are
calculated with 256 and 256$\times$256 cells respectively. 
Output is at $t$=0.1 for the
reference solution and 1D result, and at $t=0.1/\sqrt{2}$ for 2D
results with shock angle of $\alpha=45^{\circ}$. The parallel
component of the magnetic fields ($B_\parallel$) is conserved with an
accuracy of machine round-off errors.}  
\end{center}
\end{figure}

For this example, only the total energy equation is solved. The
modified entropy equation and internal energy equation are not
used. Fig. \ref{fig1} shows the results at the final time. The
parallel component of the magnetic field is preserved very well by our
CT scheme, which can preserve the divergence-free condition
($\nabla\cdot {\bf B} = 0$) to machine round-off error. 

\subsection{One-dimensional MHD Caustics}

This example is taken from Ryu et al. (1993). It was proposed as a
test for hydro cosmology code to handle the thermal energy correctly.
The formation of 1D caustics has been calculated from the initial
sinusoidal velocity field along the x-direction with the peak value
$1/2\pi$. The initial density and pressure has been set to be uniform
with $\rho=1$ and $p=10^{-10}$. Hence the initial peak velocity
corresponds to a Mach number of 1.2$\times$$10^4$. In the simulations,
the pressure can easily become negative for this type of high Mach
flow.

\begin{figure}[htbp] 
\begin{center}
\includegraphics[width=0.45\textwidth]{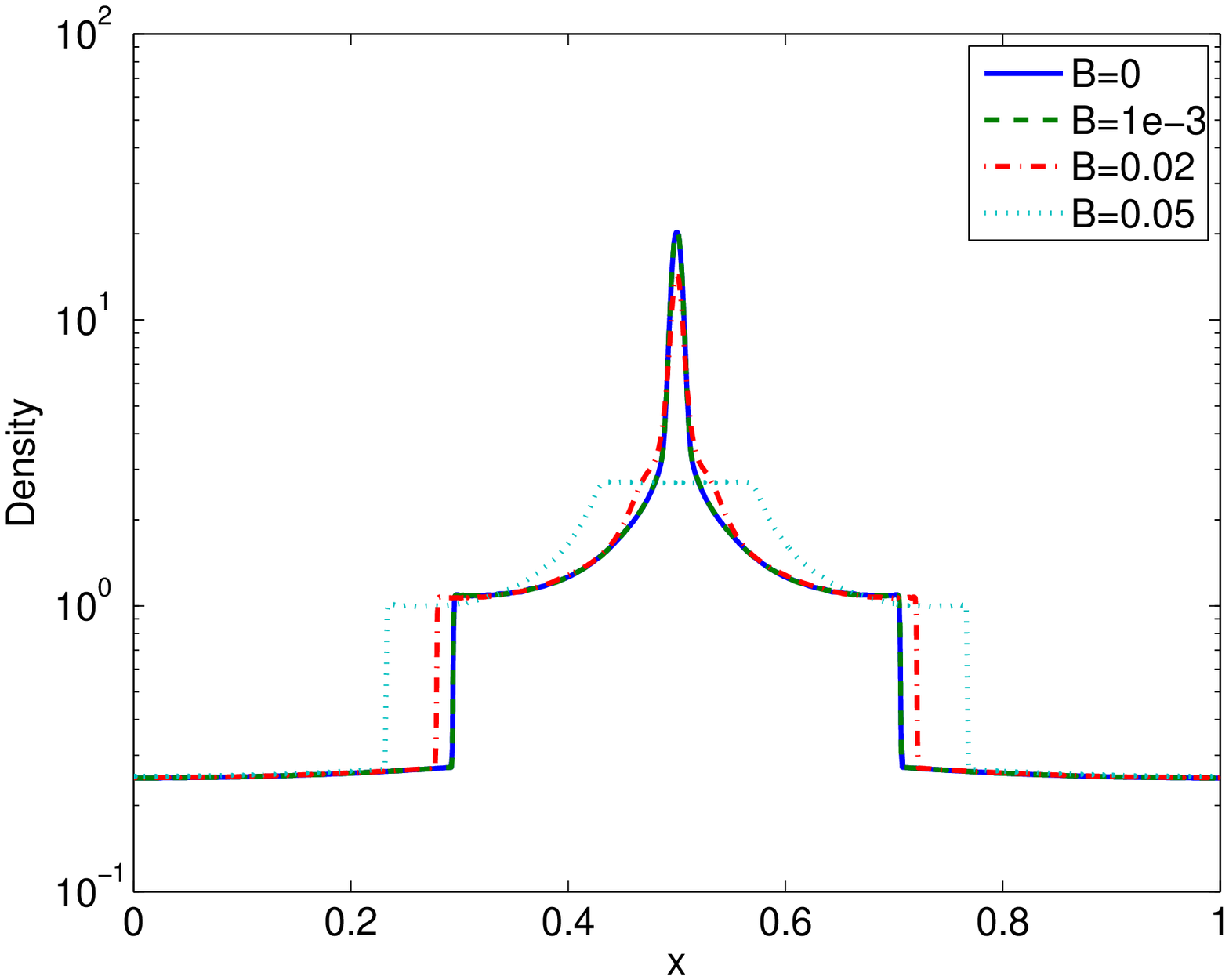}
\includegraphics[width=0.45\textwidth]{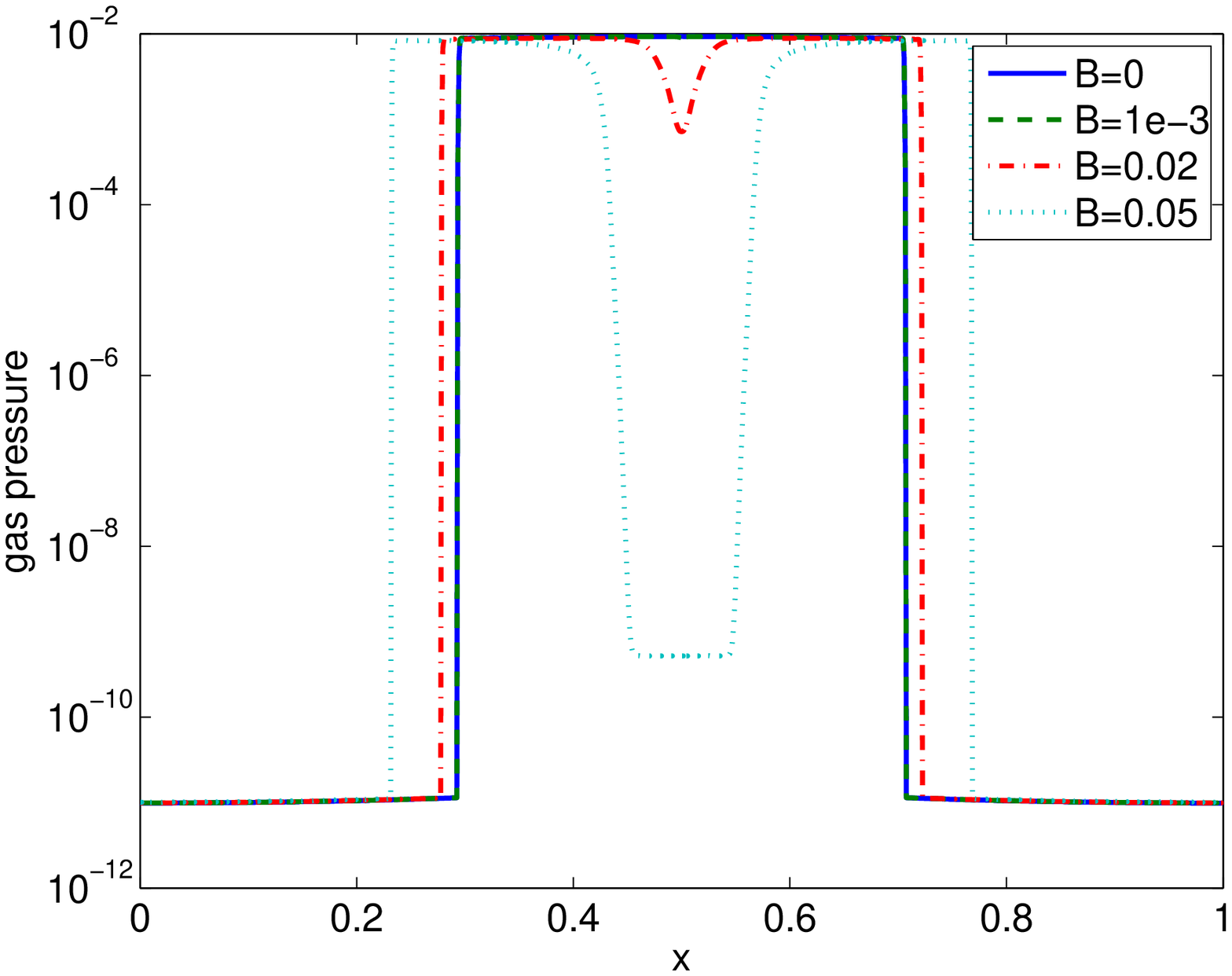}
\includegraphics[width=0.45\textwidth]{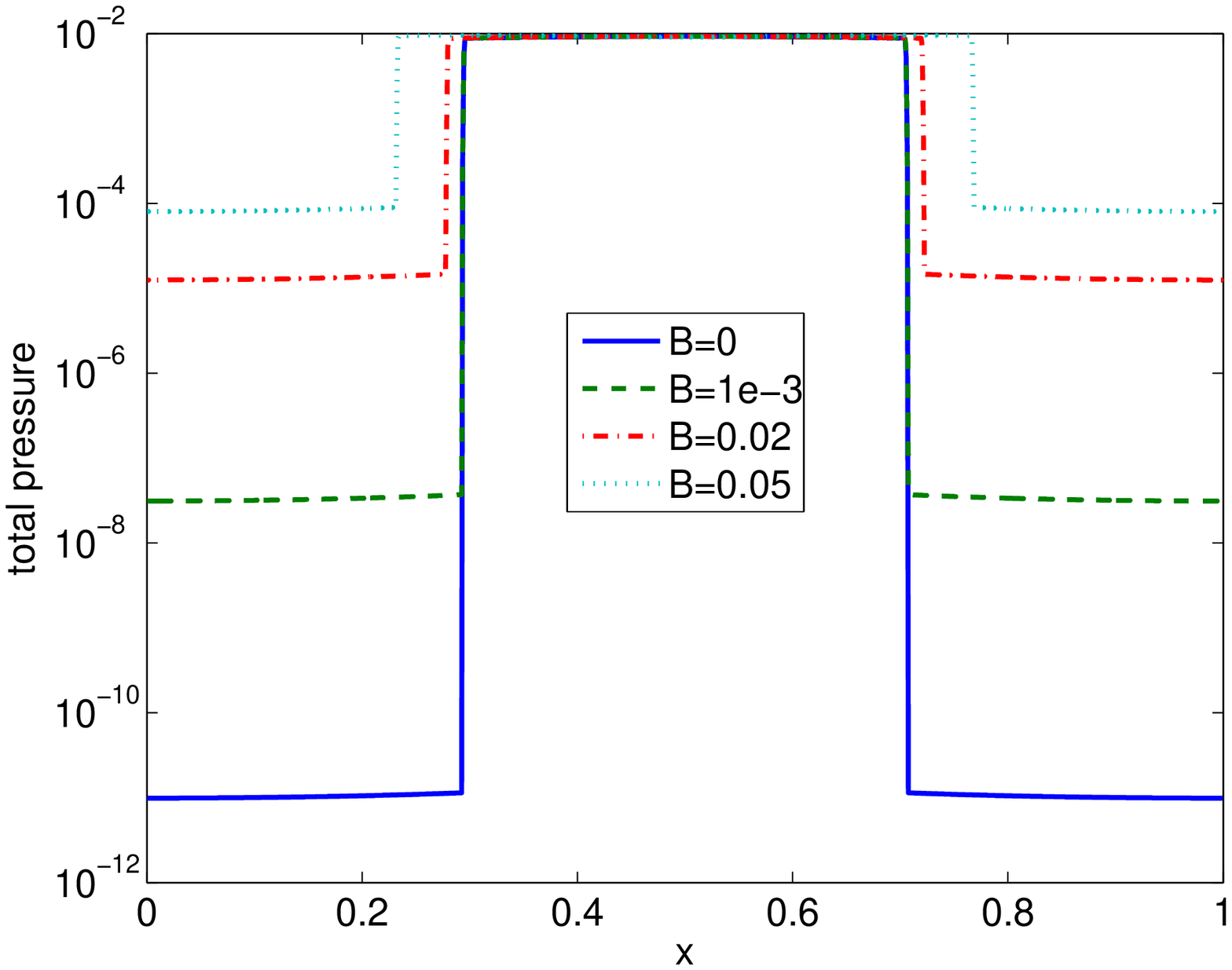} 
\includegraphics[width=0.45\textwidth]{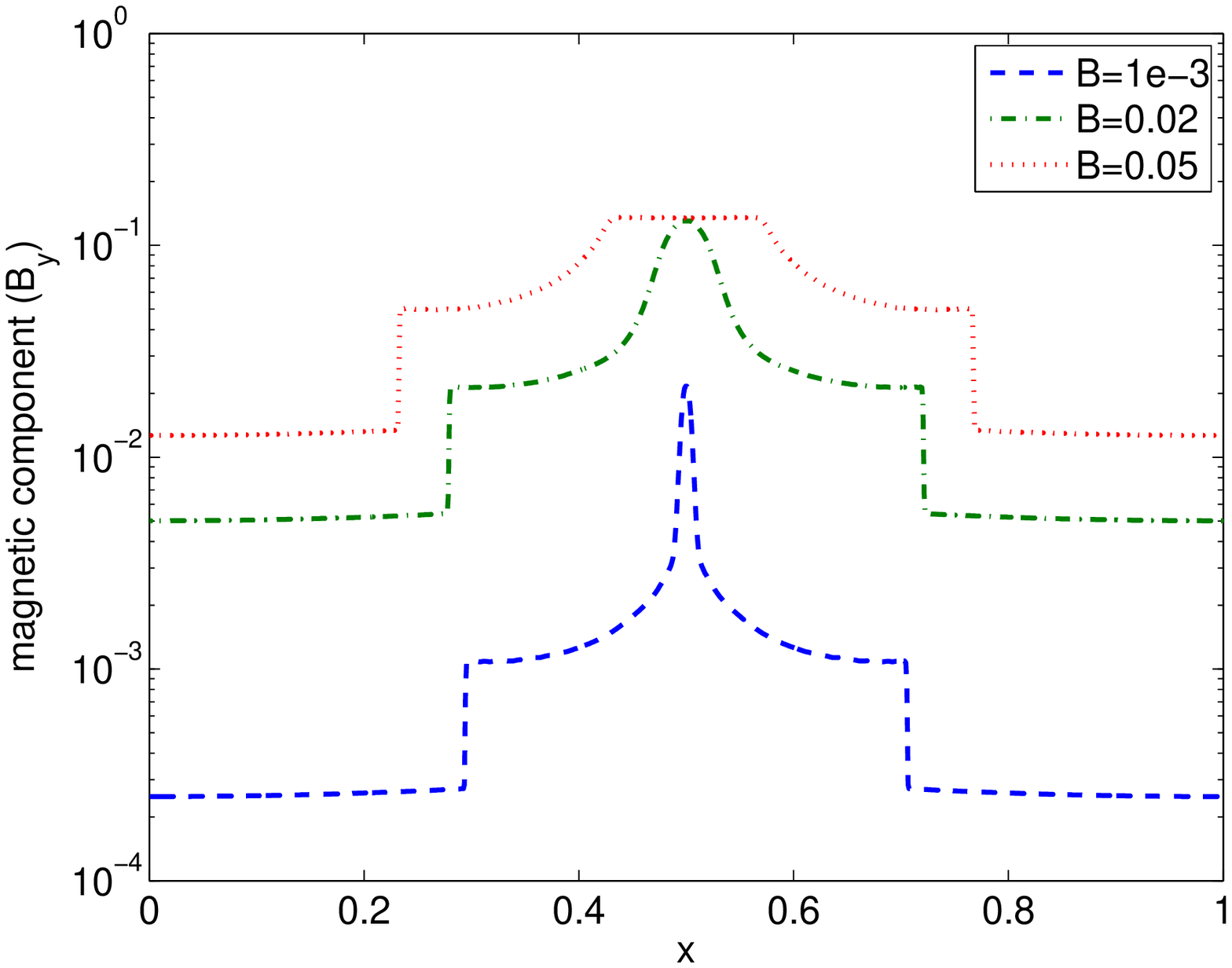} 
\caption{\label{fig2}Plots of density, gas pressure, total pressure, and
  magnetic component ($B_y$) of 1D MHD caustics test at $t=3$. The modified
  entropy equation is solved.} 
\end{center}
\end{figure}

We test this problem with and without magnetic fields.  If the
magnetic fields are present, we set $B_x=B_z=0$, and only $B_y$ is
nonzero. Different values of $B_y$ are tested.  Fig. \ref{fig2} shows
the results at $t=3$ with 1024 cells. For this problem, the total
energy version of the code will give a wrong pressure floor value,
which is not shown here. The results shown in Fig. \ref{fig2} are
calculated using the combination of the total energy version with a
modified entropy equation. It is clear that the strong magnetic fields
($B$=0.05) can suppress the density peak, while relatively weak
magnetic fields ($B$=1E-3) have little impact on the density and
pressure profile.

We also solve the internal energy equation to follow the adiabatic
changes correctly in the pre-shock region. Figure \ref{fig3} shows the
comparison between the modified entropy version and the internal
energy version for $B_y=0.05$ case. The results are almost
identical. Figure \ref{fig3} also shows the comparison with different
Riemann solvers. Apparently the HLL solver is very diffusive in the
post-shock region so that it cannot recover the gas pressure correctly
in the middle region.

\begin{figure}[htbp]
\begin{center}
\caption{\label{fig3}Comparison of different approaches (left) and
  Riemann solvers (right) with the plots of gas pressure of 1D MHD caustics
  test at $t=3$. Initial $B_y=0.05$. } 
\end{center}
\end{figure}

\subsection{One-dimensional Cosmological Pancake Collapse}

This example is also taken from Ryu et al. (1993). We set up the
problem as a 1D pancake (i.e., the Zel'dovich pancake problem) in a
purely baryonic Universe with $\Omega=1$ and $h=0.5$ in the comoving
coordinates. The MHD solver with self-gravity module is tested.
Initially, at $a_i=1$, which corresponds to $z_i=20$ in this test, a
sinusoidal velocity field with the peak value $0.65/(1+z_i)$ in the
normalized units has been imposed in a box with the comoving size
$64h^{-1}$Mpc. The baryonic density and pressure have been set to be
uniform with $(\rho,p)=(1,6.2\times10^{-8})$ in the normalized
units. The calculation has been done with a different number of cells,
with and without magnetic fields. When the magnetic fields are
present, we use only the $B_y$ component, and set $B_x=B_z=0$.

\begin{figure}[htbp]
\begin{center}
\includegraphics[width=0.8\textwidth,height=0.28\textheight]{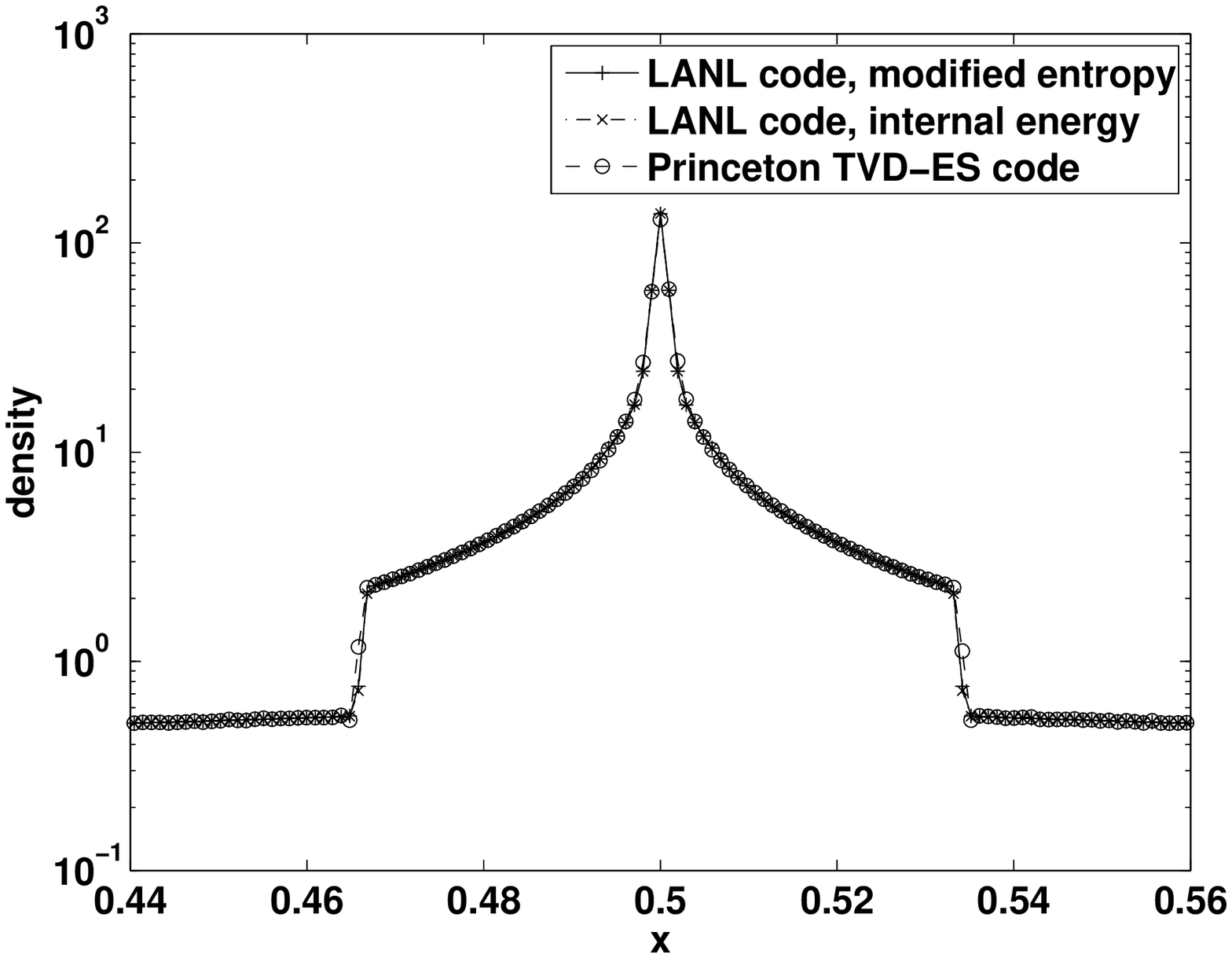}
\includegraphics[width=0.8\textwidth,height=0.28\textheight]{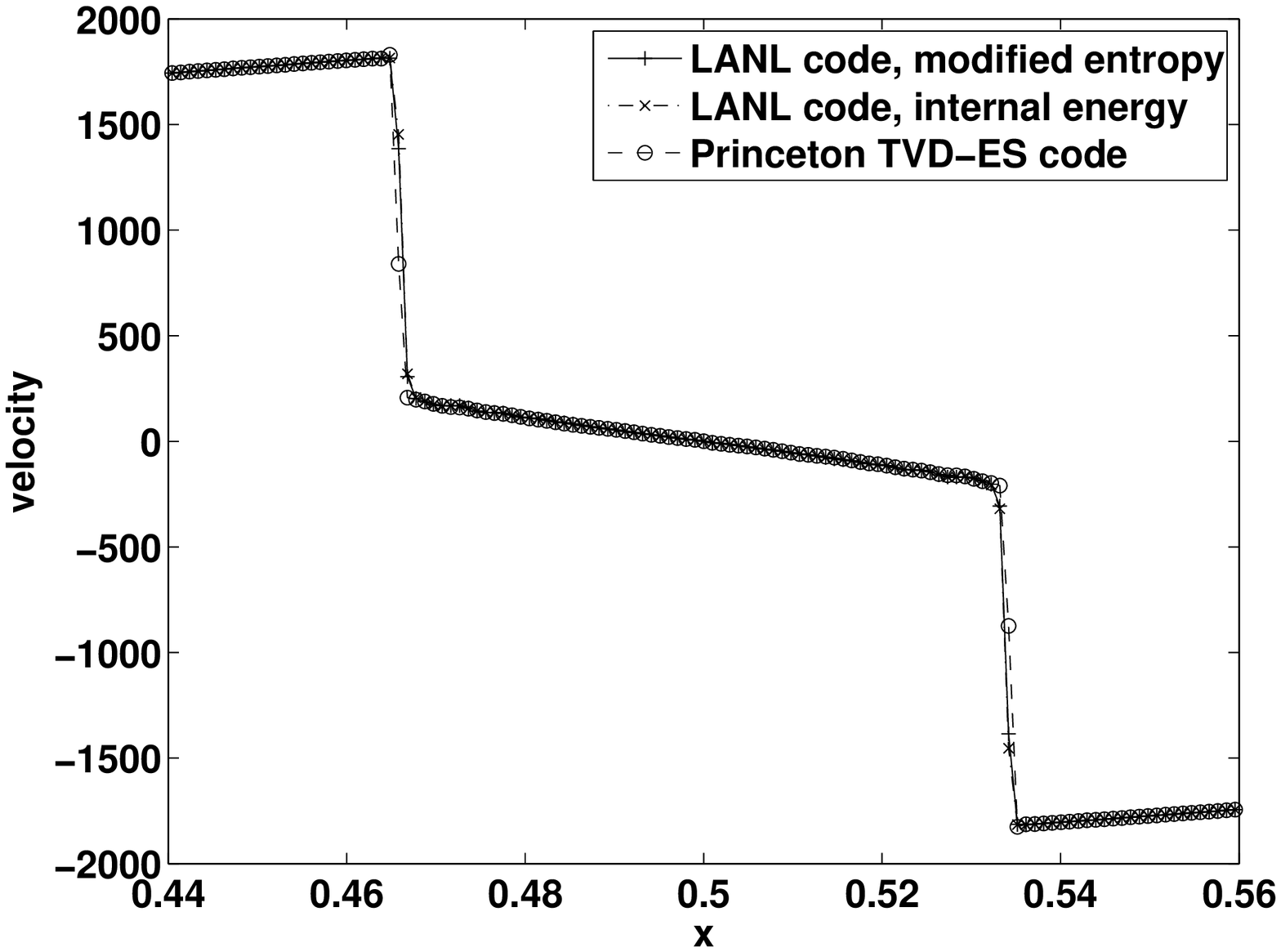}
\includegraphics[width=0.8\textwidth,height=0.28\textheight]{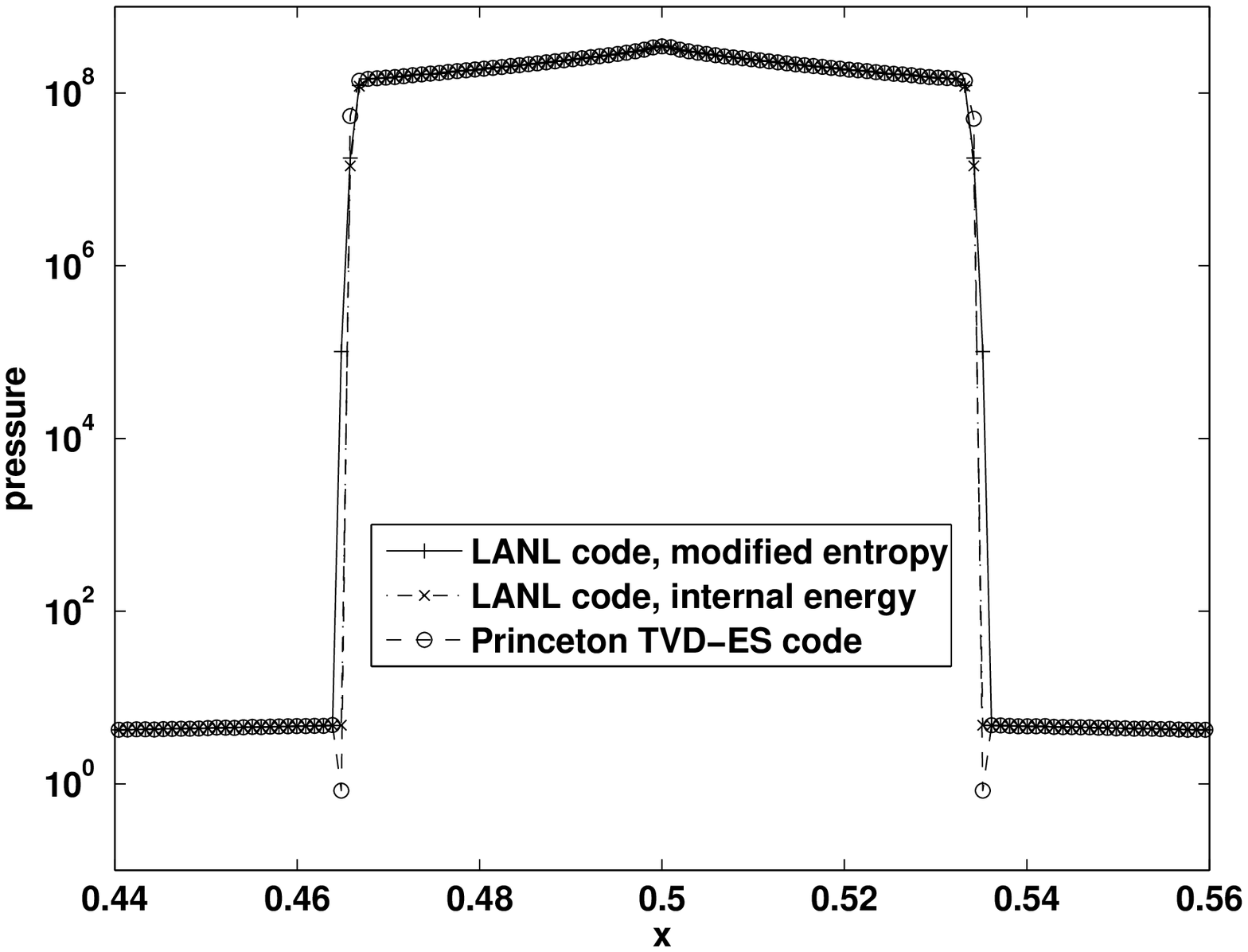}

\caption{\label{fig4} One-dimensional gravitational collapse of a
  pancake in the comoving coordinates with different codes. The number
of cells is 1024. No magnetic fields are applied ($B=0$).} 
\end{center}
\end{figure}

Figure \ref{fig4} shows the results with $B_y=0$, i.e., without the
magnetic fields. The results for both the modified entropy version and
internal energy version are shown. For comparison, we also show the
results of TVD-ES code (Ryu et al. (1993)),
running with the mass-diffusive correction.

\begin{figure}[htbp]
\begin{center}
\includegraphics[width=0.45\textwidth]{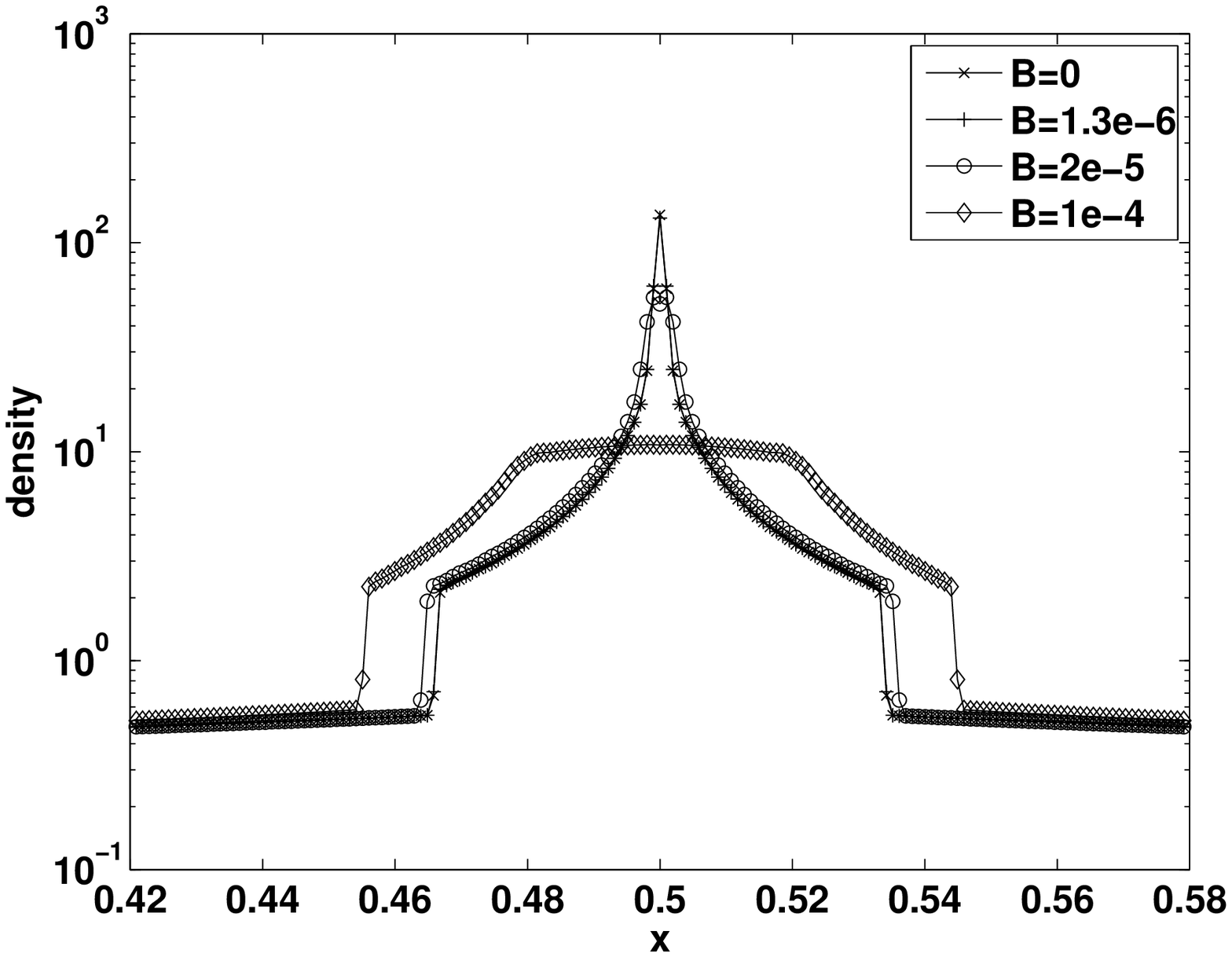}
\includegraphics[width=0.45\textwidth]{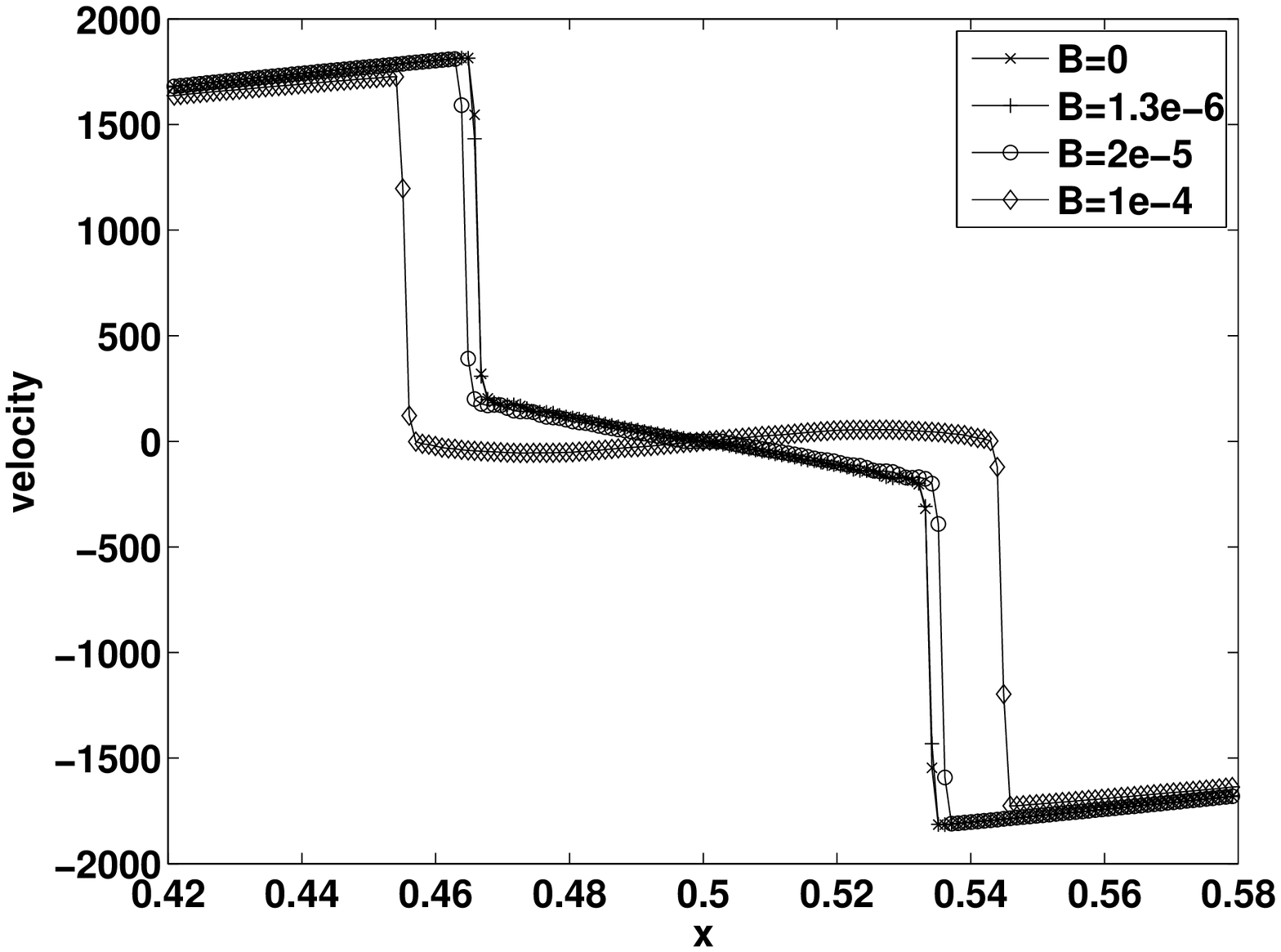}
\includegraphics[width=0.45\textwidth]{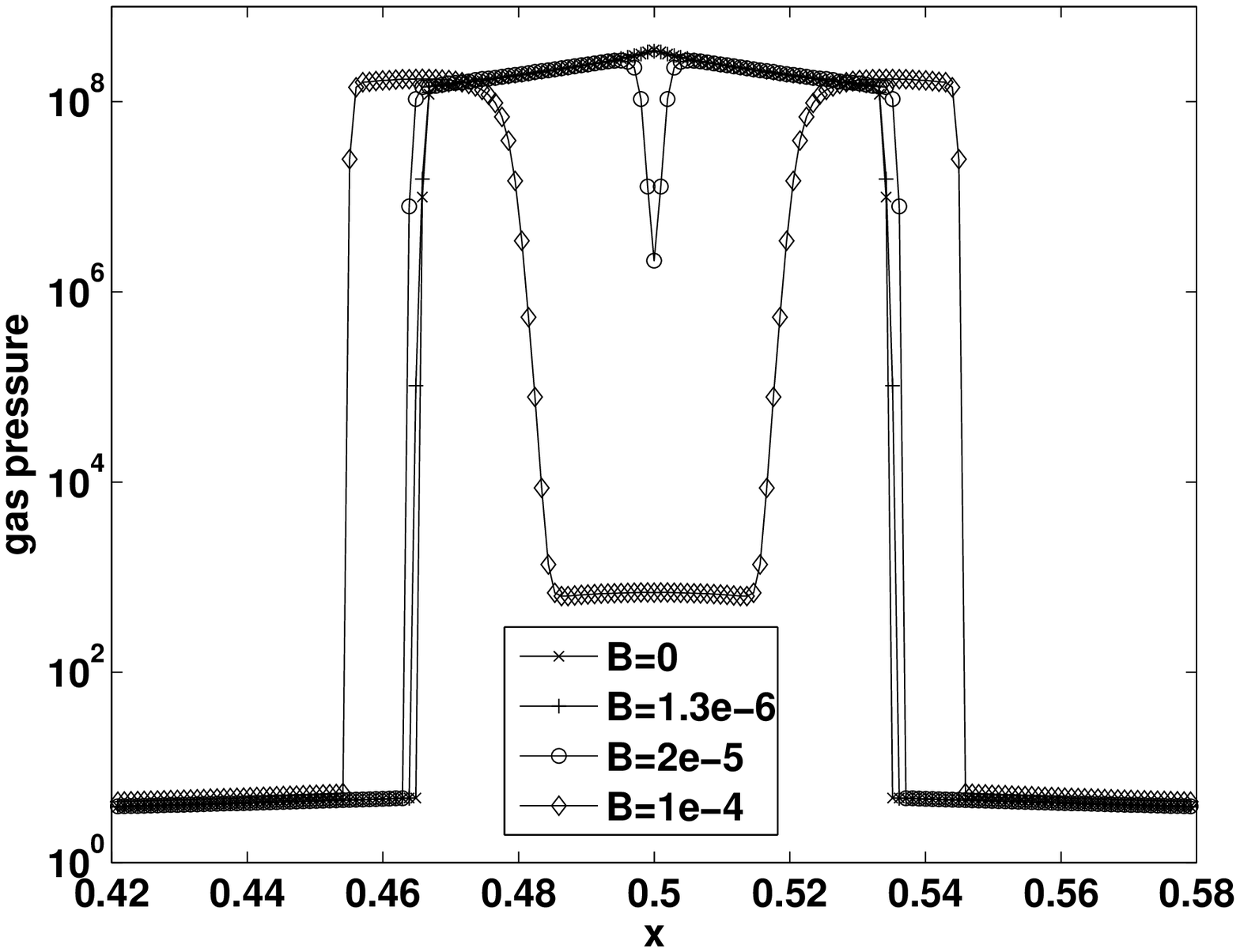}
\includegraphics[width=0.45\textwidth]{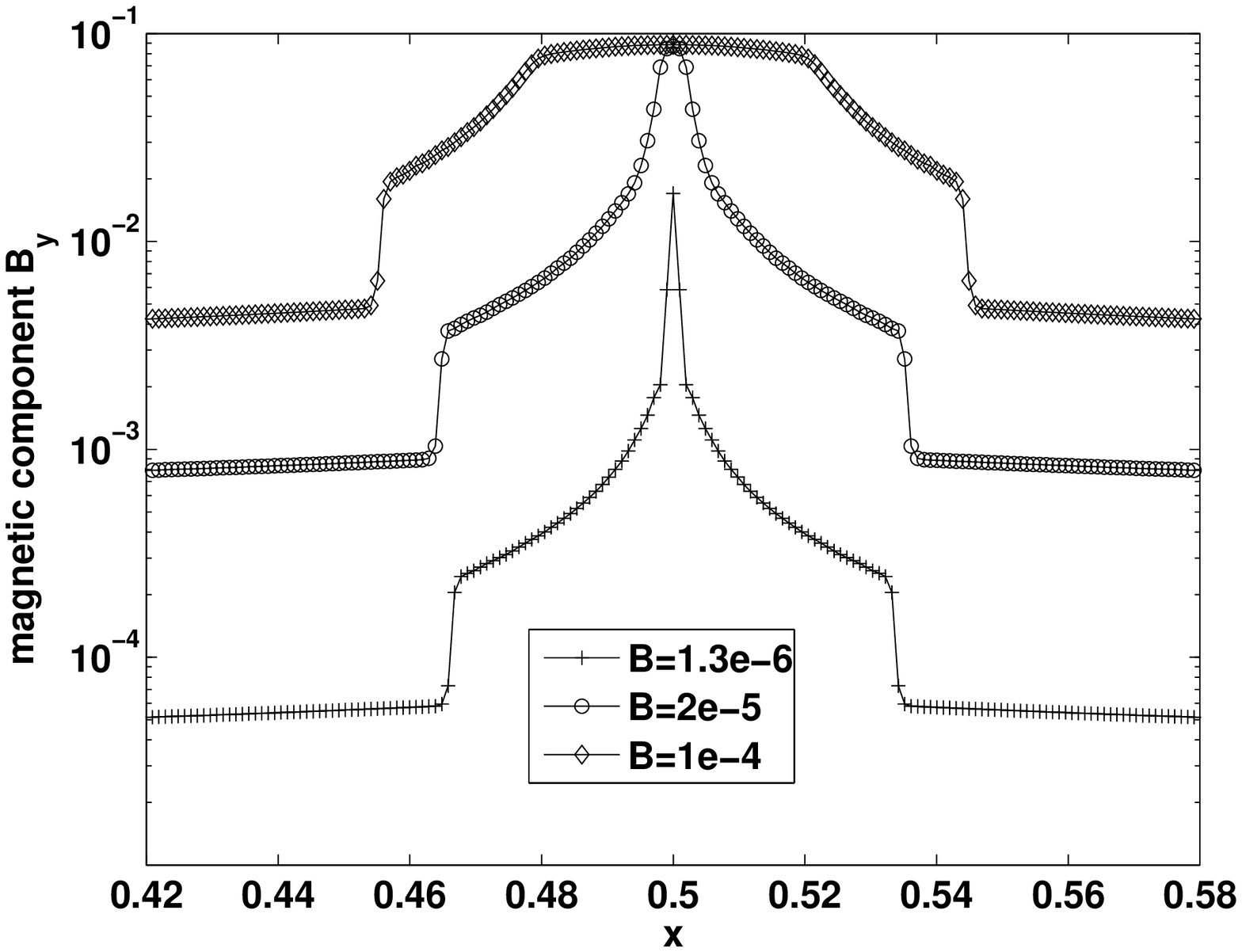}
\caption{\label{fig5} One-dimensional gravitational collapse of a
  pancake in the comoving coordinates with different magnitude of
  magnetic fields. The number of cells is 1024. } 
\end{center}
\end{figure}

Figure \ref{fig5} shows the results with magnetic fields. Different
magnetic fields have different impact on the shock location and
the peaks of density and magnetic fields. With a small $B_y=1.3e-6$,
the magnetic fields have little impact on the shock and the density
profile. The magnetic fields collapse in the same way
as the density field does. As we increase the magnitude of the magnetic
fields, the density peak in the post-shock region and shock location
show large changes, while the profile in the pre-shock region has
little change.

\begin{figure}[htbp]
\begin{center}
\includegraphics[width=0.45\textwidth]{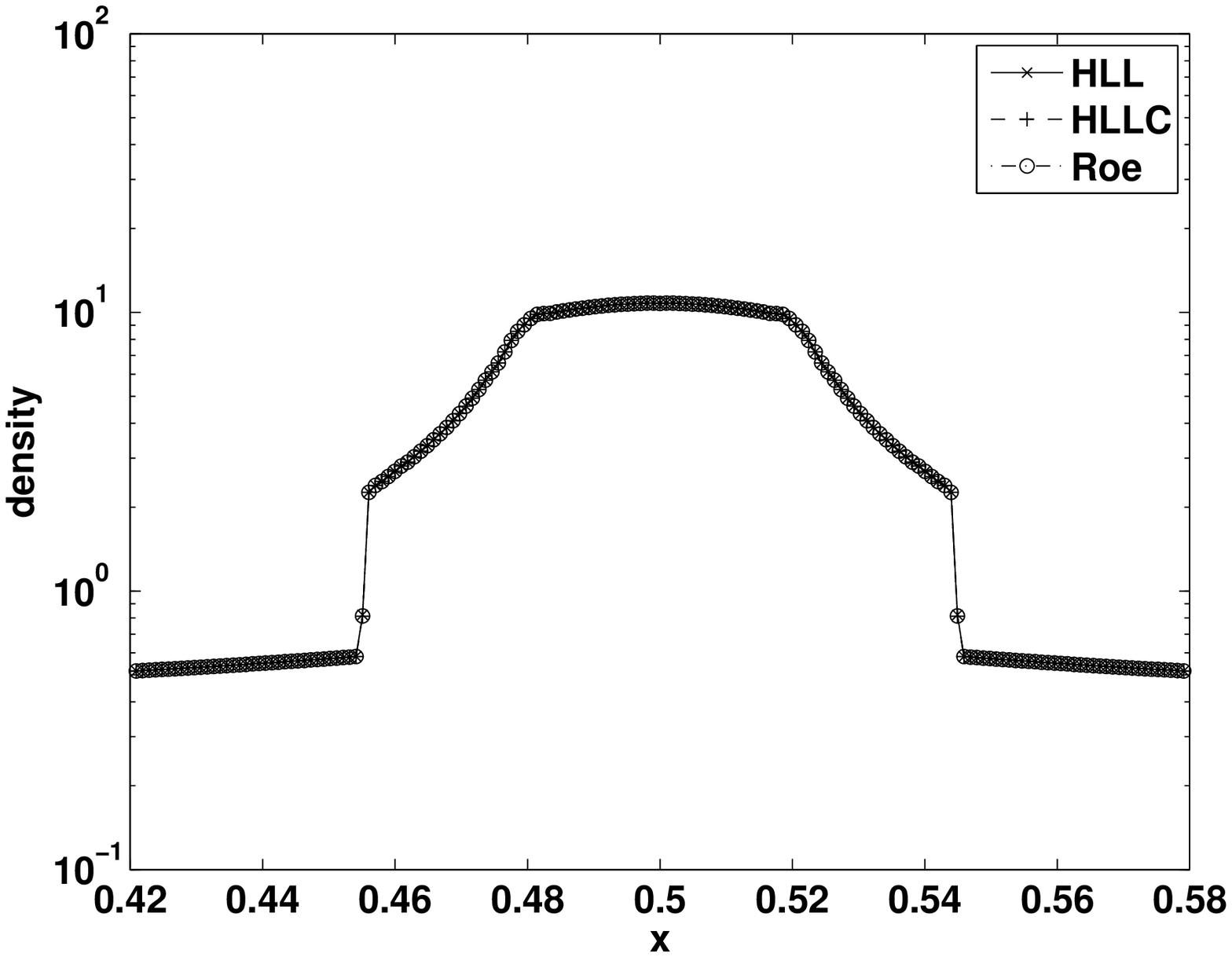}
\includegraphics[width=0.45\textwidth]{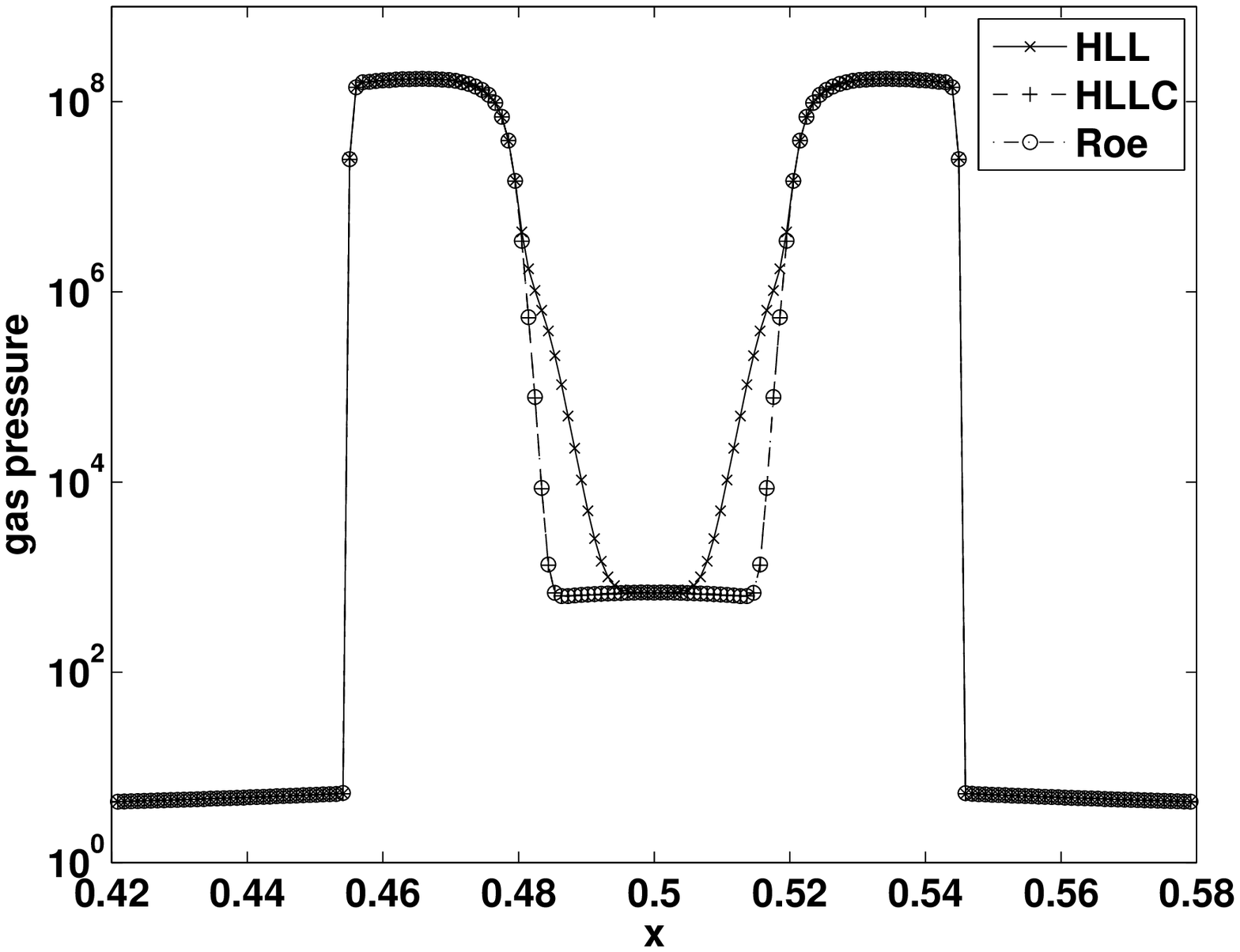}
\caption{\label{fig6} One-dimensional gravitational collapse of a
  pancake in the comoving coordinates with different Riemann
  solvers. The number of cells is 1024. $B_y=10^{-4}$.} 
\end{center}
\end{figure}

Figure \ref{fig6} shows the comparison for different Riemann solvers
applied to the modified entropy equation.  The results for HLLC and
the more sophisticated Roe's solver are almost identical, while the
result for HLL solver is not so good at the post-shock region.

\subsection{Comparison with Other Codes for 3D Adiabatic CDM Universe}

We have carried out a comparison of our cosmological MHD code with the
original TVD-ES code of Ryu et al. (1993). We computed the adiabatic
evolution of a purely baryonic Universe but with an initial CDM power
spectrum with the following parameters: $\Omega = \Omega_b =
1,~h=0.5,$, $n=1$ and $\sigma_8=1$, and the size of the computational
cube is $L=64 h^{-1}{\rm Mpc}$.  We use the Bardeen \etal (1986)
transfer function to calculate the power spectrum of the initial
density fluctuations.  This test problem is identical to that of Kang
\etal (1994) except that the random numbers used to generate the
initial condition are different.  The Universe is evolved from $z=30$
to $z=0$.  We use $256^3$ cells for each simulation performed.  The
comparisons are made at the final epoch, $z=0$.

\begin{figure}[htbp]
\begin{center}
\includegraphics[width=0.8\textwidth]{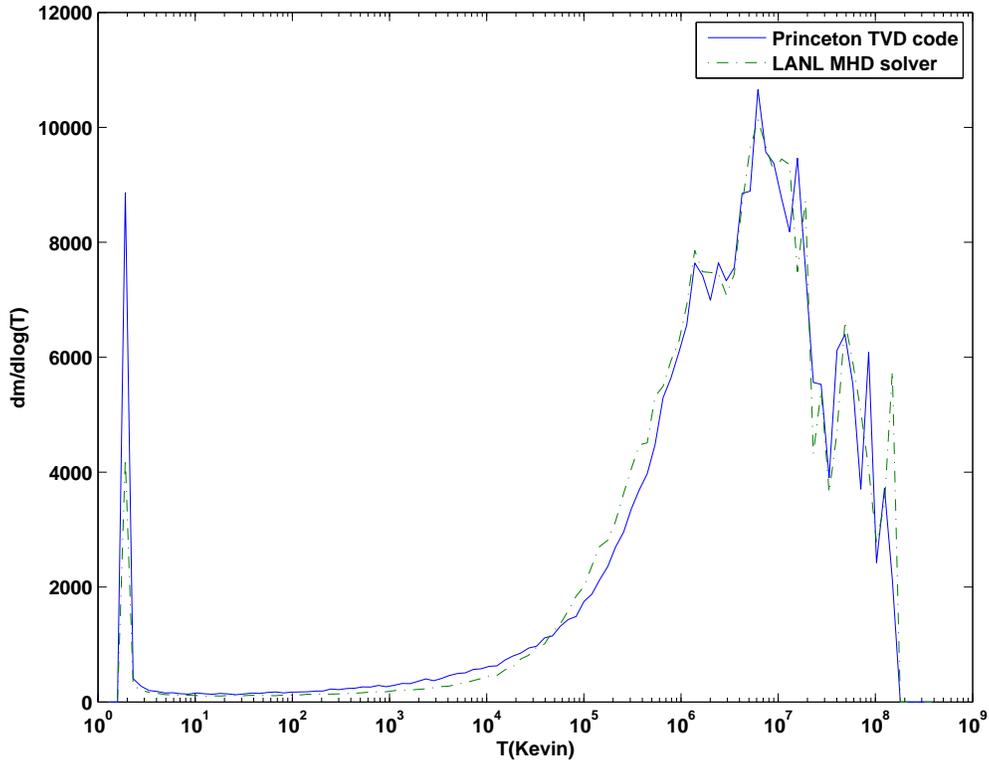}
\caption{\label{fig7} Mass-weighted temperature histogram at present
  (redshift z=0) for the 3D simulations of a purely baryonic adiabatic
  Universe.}
\end{center}
\end{figure}

Figure \ref{fig7} shows a comparison of a mass-weighted histogram of
the baryonic temperature at $z=0$ without magnetic fields between the
original TVD-ES code and CosmoMHD. The mass-weighted histogram is
calculated by using the temperature as $x$-axis and total mass in each
temperature bin as the $y$-axis. They are in a good agreeement.

We have also performed another run with initial magnetic fields,
$B_x=B_z=0$, $B_y$=2.5E-9 Gauss (which corresponds to 4.3176E-7 in
code unit). Figure \ref{fig8} shows a slice of the density and
magnetic energy at $z=0$.  We see that the distributions of density
and magnetic field are strongly correlated, as expected.  To
demonstrate the impact of the CT method, we set the initial $B_y$=1E-5
Gauss, which corresponds to 0.0017 in code unit.  Figure \ref{fig9}
shows the comparison of the divergence of the magnetic fields,
averaged over the entire box, as a function of redshift.  It is clear
that without the CT method, the divergence of the magnetic fields
grows in tandem with the growth of the magnetic field strength, and
simulation results would hardly be meaningful.  Since single-precision
is used in our whole simulation, the divergence from the CT method is
close to the round-off error.  The temperature plots are shown in
Fig. \ref{fig10}. The plasma beta becomes very small ($10^{-6}$) in
many regions with these large magnetic fields. The dual-formulation
must be used to track the internal energy accurately.

\begin{figure}[htbp]
\begin{center}
\caption{\label{fig8} Density and magnetic energy at present for the
  3D simulations of a purely baryonic adiabatic Universe. Left: log
  density plot at z=0.5 slice. Right: log magnetic energy plot at
  z=0.5 slice. Initial magnetic fields are $B_x=B_z=0$ and
  $B_y$=4.3176E-7.}
\end{center}
\end{figure}

\begin{figure}[htbp]
\begin{center}
\includegraphics[width=0.8\textwidth]{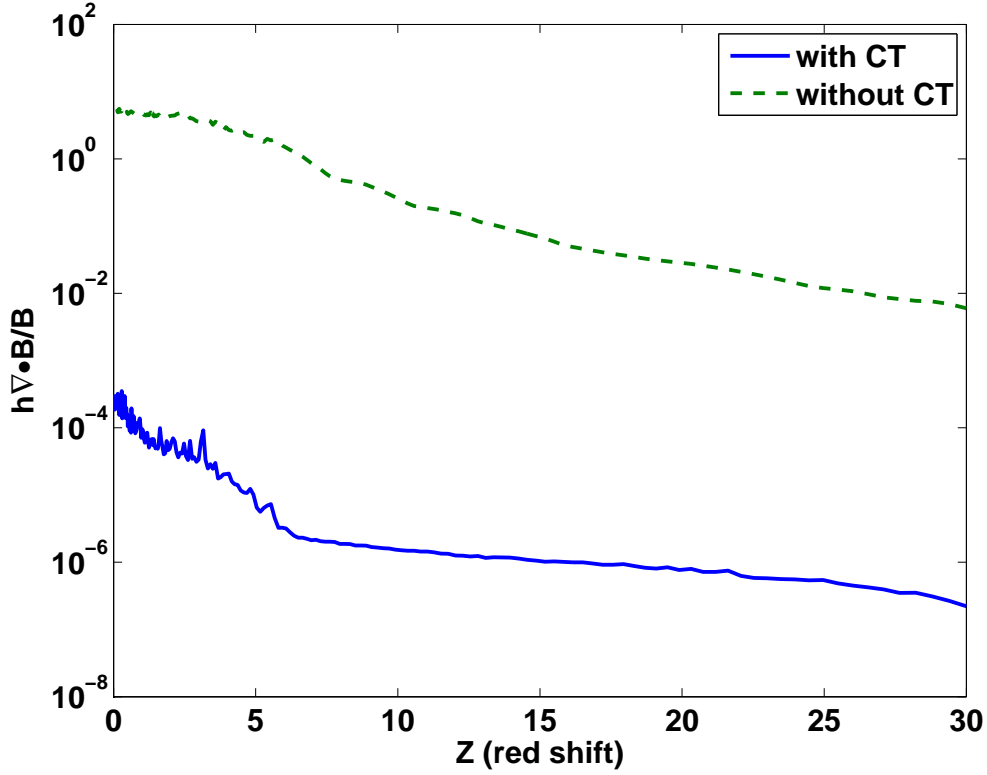}
\caption{
  \label{fig9} The scaled divergence ($h\nabla\cdot {\bf B}/B$) of the magnetic fields with and without
  constraint transport (CT) for the three-dimensional simulations of a
  purely baryonic adiabatic Universe during the whole time
  history. The initial $B_y$ at $Z=30$ is 0.0017.  }
\end{center}
\end{figure}

\begin{figure}[htbp]
\begin{center}
\caption{\label{fig10} Temperature plots of a slice at present for the
  3D simulations of a purely baryonic
  adiabatic Universe. Left: log temperature plot without using CT at z=0.5
  slice. Right: log temperature plot with CT at z=0.5 slice. Initial magnetic
  fields are $B_x=B_z=0$ and $B_y$=0.0017.} 
\end{center}
\end{figure}

\section{\label{sec:conclu}Discussion and Conclusions}

We have developed and tested a modern cosmological MHD code
(CosmoMHD).  In the CosmoMHD code, five sets of equations are evolved
simultaneously: (1) the ideal MHD equations for magneto-gasdynamics,
(2) rate equations for multiple species of different ionizational
states (including hydrogen, helium and oxygen), (3) the Vlasov
equation for dynamics of collisionless particles (including dark
matter particles and stellar particles), (4) the Poisson's equation
for obtaining the gravitational potential field and (5) the equation
governing the evolution of the intergalactic ionizing radiation field,
all in cosmological comoving coordinates.  The MHD solver consists of
several high-resolution schemes with finite-volume and
finite-difference methods.  The divergence-free condition of the
magnetic field is preserved to an accuracy due just to round-off
errors.  Additional implemented physical processes include: (1)
detailed cooling and heating processes due to all the principal line
and continuum processes for a plasma of primordial composition as well
as cooling and heating due to metals, (2) star formation processes and
feedback processes from star formation including UV radiation and
galactic superwinds, and (3) formation of supermassive black holes and
associated feedback processes including radio jets/lobes.

The CosmoMHD code developed here may be applied to a variety of
astrophysical/cosmological problems with or without self-gravity.  We
intend to study the formation and evolution of galaxies, clusters of
galaxies and the intergalactic medium using this accurate code.  Most
importantly, we would like to investigate the role played by the AGN
feedback in the form of powerful radio jets/lobes in regulating
structure formation on scales from the cores of clusters of galaxies
($\sim 10$kpc) through giant lobes ($\sim 100$kpc$-1$Mpc) to large
scales ($>1$Mpc).  It seems prudent to check if many fundamental
cosmological applications, such as weak gravitational lensing,
formation and evolution of X-ray/SZ clusters and matter power spectrum
with baryonic oscillation features, may be affected by this feedback
effect.  A thorough understanding of AGN feedback may be necessary to
reduce systematic errors to levels that are commensurate with
statistical errors that many of these important observations are
expected to be able to achieve.

\acknowledgments 

We thank Eric Johnson for coding a prescription to form initial
realistic MHD radio jets associated with SMBH formation. This work was
carried out under the auspices of the National Nuclear Security
Administration of the U.S. Department of Energy at Los Alamos National
Laboratory under Contract No. DE-AC52-06NA25396, and is supported by
the Laboratory Directed Research and Development programs at LANL. It
is also supported in part by grants AST-0407176, AST-0507521,
NNG05GK10G and NNG06GI09G.

\end{document}